\documentclass[11pt]{article}
\pdfoutput=1
\usepackage{times}
\usepackage{helvet}
\usepackage{courier}
\usepackage[margin=1in]{geometry}
\usepackage[numbers,sort&compress]{natbib}
\usepackage{hyperref}
\usepackage{graphicx}
\usepackage{amsmath}
\usepackage{amssymb}
\usepackage{amsthm}
\usepackage[nameinlink]{cleveref}
\usepackage{thmtools}
\usepackage{xcolor}
\usepackage[normalem]{ulem}
\usepackage{authblk}

\newtheorem{example}{Example}

\begin{document}

\title{AI of the People, by the People, for the People: A Social Choice Approach to Collective Control of Artificial Intelligence}

\author[1]{Paul Anton Bachmann}
\author[2]{Niclas Boehmer}
\author[3]{Lukas Daniel Klausner}
\author[3]{Martin Lackner}

\affil[1]{Institute of Logic and Computation, TU Wien, Vienna, Austria}
\affil[2]{Hasso Plattner Institute, University of Potsdam, Potsdam, Germany}
\affil[3]{Center for Artificial Intelligence, University of Applied Sciences St.\ P\"olten, St.\ P\"olten, Austria}

\date{}

\maketitle

\begin{abstract}
With the growing adoption of AI systems, reasoning about how society can exert control over AI becomes an increasingly urgent problem. Existing work on democratic control largely focuses on macro-level governance. In contrast, we propose a new approach grounded in social choice theory, which we term \emph{collective control of artificial intelligence}. We argue that collective input can and should be incorporated at multiple points across the ML development pipeline, from data collection through objective design to alignment. We further demonstrate that social choice provides a well-suited modelling language for the treatment of collective input across all stages and that its axiomatic methodology yields principled criteria for evaluating various control mechanisms. Overall, our conceptual contribution provides a mathematically grounded framework to implement and analyse collective control of AI systems.
\end{abstract}


\section{Introduction}\label{sec:introduction}
The growing adoption of machine learning (ML) and artificial intelligence (AI), and in particular the widespread deployment of large language models (LLMs) and other foundation models~\cite{Shah2023, Yan2024, Makridakis2023}, has intensified questions around how societies can and should exert control over AI systems.
In fact, the seemingly ever-growing range of domains and use cases in which AI systems are applied has brought issues of participation, power and democratic legitimacy to the forefront of societal and scientific debates.
Accordingly, calls for collective governance of AI have emerged across disciplines, from computer science~\cite{Bahrami2025, BogiatzisGibbons2024, Kallina2025, Suresh2024} through economics~\cite{Kasy2024,Kasy2025}, philosophy~\cite{Sangiovanni2019} and political science~\cite{Prainsack2025} to civil society organisations~\cite{WGAI2024}. These concerns also connect naturally to established accounts of trustworthy AI, including the ethical guidelines of the European Commission's High-Level Expert Group on Artificial Intelligence~\cite{HLEG2019}.
More generally, societal control of AI systems can be understood as an immediate implication of the ethical principle of respect for human autonomy and the associated requirement of ensuring human agency and oversight.
In this paper, we therefore adopt the premise that some form of public oversight and regulation is appropriate and, indeed, necessary, while pointing to recent literature~\cite{Mesko2023, Goanta2023, Berengueres2024, Hacker2023} for a more comprehensive debate on the merits and limits of regulation of AI.

Existing work on democratic control of AI has largely focused on macro-level governance and institutional design (e.\,g.\ \cite{Aytaç2024,BogiatzisGibbons2024,Furendal2025}; see \Cref{sec:background} for a more extensive discussion of previous work).
Complementing this line of research, we propose a micro-to-meso level framework for what we term \emph{collective control of AI}: incorporating collective decision-making mechanisms in the AI development and deployment process.
We thus position collective control of AI as the concrete technical and socio-technical counterpart to regulatory control of AI. This is in stark contrast to the current reality of AI development and deployment, where consequential design choices are typically made centrally by a small number of decision-makers.

Our central claim is that collective decisions arise repeatedly across the ML development pipeline and that these decisions and their consequences can be analysed in a unified way using social choice theory~\cite{list-sc,brandt2016handbook}. Social choice theory studies collective decision-making problems, in which possibly conflicting preferences of voters over alternatives need to be aggregated into a compromise solution.
We use this social choice lens to make the collective decision dimension of ML development explicit:
At many stages of the ML development pipeline, various stakeholders have legitimate interests over the decisions taken at this step that can be modelled as preferences over a set of possible decision options. To be able to account for such potentially conflicting preferences in the decisions taken at some stage, they must be aggregated into a collective decision, a process that usually takes place implicitly (or not at all).
We argue that the axiomatic toolbox of social choice theory provides principled criteria for evaluating the desirability of such mechanisms for collective control.

Our contributions are as follows:
\begin{itemize}
    \item In \Cref{sec:background}, we give a brief overview of the existing literature on collective or democratic control of AI systems and how our approach relates to existing ones.
    \item In \Cref{sec:mlpipeline}, we provide a schematic view of the ML development process and identify concrete decision points at which stakeholder preferences are relevant and can be incorporated. We argue that meaningful collective control requires attention to all of these decision points.
    \item We develop a formal framework for understanding collective control via social choice theory, operating at two complementary levels:
    \begin{itemize}
        \item In \Cref{sec:ai-as-scfs}, we develop a formal model of end-to-end collective control by modelling deployed AI systems as social choice functions, enabling us to apply established axiomatic criteria to evaluate how systems respond to stakeholder preferences.
        \item In \Cref{sec:approaches}, we describe a new approach for collective control in the ML development process by connecting each decision point in ML development to established social choice problems (such as multi-winner voting, portioning and epistemic voting), showing how existing mechanisms and axioms can guide the design of fair and efficient collective control procedures for these decision points.
    \end{itemize}
\end{itemize}

\noindent The framework we propose is intended as a first step towards bringing together social choice theory and research on the control of AI systems. Taken together, \Cref{sec:ai-as-scfs} and \Cref{sec:approaches} demonstrate that social choice can provide both a unifying formalism and principled guidance for collective control both at the level of end-to-end deployed behaviour and at specific decision points across the ML development pipeline. Beyond providing ideas for new mechanisms, our work takes a first step towards formally grounded and actionable evaluation methods for assessing the degree of responsiveness to collective preferences realised in existing systems and processes.

\section{Background and Related Work}\label{sec:background}

As AI systems increasingly support, influence and even make decisions with societal consequences, various approaches to democratising this technology have been proposed. The term `democratisation of AI' is used to refer to a multitude of ideas pursuing different and possibly conflicting goals~\cite{seger2023DemocratisingAIMultiple}. In this work, we specifically address the democratisation of AI \emph{governance}, with the aim of developing principled methods to reduce unilateral decision-making and allow for democratic decisions around the development and use of AI systems.

To this end, a growing body of research has emerged at the intersection of AI alignment, democratic governance and social choice theory. Rather than converging on a single paradigm, this literature explores a range of approaches to reflecting and incorporating collective input in the design, deployment, management and control of AI systems. Our work contributes to these debates by considering the full spectrum of AI development and deployment and by framing collective control as a problem of social choice. In the following, we review related branches of research.

\textbf{Alignment and social choice.} A prominent line of work related to collective input into the training of AI systems is AI alignment, which studies how to integrate diverse human inputs into system behaviour, including approaches based on machine ethics~\cite{Anderson2007}, value learning and reinforcement learning from human feedback~\cite{Casper2023RLHF}. Within this area, several contributions draw explicitly on social choice theory to formalise the aggregation of diverse human preferences, treating alignment as a collective decision problem~\cite{Conitzer2024,noothigattu2018voting,Anthropic2024,ge2024axioms}.
While similar in spirit, our approach differs in scope: we do not focus solely on fine-tuning, but consider collective control across the broader development and deployment pipeline. Related work has explored learning aggregation rules directly from data and replacing `classical' social choice functions with learned functions based on neural networks trained on preference profiles~\cite{hornischer2025learning, burka2022voting, matone2024deepvoting, procaccia2009learnability, xia2013designing, anil2021learning, mohsin2022learning}. This research direction aims to discover new social choice functions with unmatched desirable properties. In contrast, we treat AI systems themselves as social choice functions and ask how collective decision-making can be justified and structured throughout their development.

\textbf{Democratic alignment and control.} Beyond technical challenges in AI alignment, recent work emphasises that democratic control of AI raises institutional and procedural questions that cannot be reduced to aggregating feedback~\cite{Aytaç2024,BogiatzisGibbons2024,Furendal2025,Lowe2025,Zimmermann2025,ZhiXuan2025}. These contributions stress, among other themes, ongoing oversight, legitimacy and representation, and argue that alignment must go beyond preference aggregation to accommodate deliberation, moral disagreement and collective values that are malleable, context-dependent and evolving. Building on these insights, we further distinguish collective control from alignment: While alignment is an important component, collective control also concerns among others who participates at which stages, which design decisions are subject to collective input and how such input is translated into influence over outcomes.

\textbf{Participation, data and power.} Work on participation in AI systems has identified the distributed structure of data generation as paralleling participatory design and has partly attributed the strength of modern AI systems to the involvement of many data-generating ``co-designers''~\cite{sloane22participation}. At the same time, this line of work highlights structural limitations: Data contributors often do not recognise their contributions (e.\,g.\ CAPTCHA challenges, location data or online reviews) as valuable or even as work and they typically lack recognition, compensation and meaningful influence over the terms of participation or system design. We agree with this critique and view it as a broader issue beyond implicit participation through data collection alone. We argue that collective input should extend beyond data provision to include the infrastructure, objectives and design choices that shape the full AI system pipeline.

\textbf{Frameworks for democratic AI governance.} Recent frameworks strive to systematise different degrees of democratic involvement in AI systems: The \emph{Democracy Levels Framework} distinguishes several levels of participation from minimal consultation to substantive collective control~\cite{Ovadya2025}. Complementary initiatives, such as deliberative mechanisms (``alignment assemblies'') as for instance tested by the \emph{Collective Intelligence Project}~\cite{CIP2023}, explore how processes of collective governance can be implemented in practice.
Further, similar to our work, the \emph{Collective Governance for AI} framework by Metagov~\cite{metagov-cg-ai} identifies multiple intervention points across the AI development pipeline and points to concrete strategies and existing projects for incorporating collective input at each stage.
Efforts such as these provide an important backdrop for our contribution, while being more focused on practically implementable governance methods.
However, unlike these previous works, our approach is not meant to propose specific governance mechanisms, but to clarify (using principles and tools from social choice theory) where collective decisions arise in the development of AI systems and how higher levels of collective control can be described, characterised and evaluated.

\textbf{Positioning, scope and non-goals.} Prior work has already drawn several important connections between AI alignment, democratic governance and social choice theory. Our contribution complements existing research by not limiting itself to isolated mechanisms or decision points; instead, we map opportunities for collective control across the AI development pipeline. In doing so, we adopt an  axiomatic, conceptually grounded approach attentive to questions of power, participation and legitimacy. Our article focuses on normative and conceptual structures of collective control in AI systems and uses tools from social choice theory to identify and analyse key decision points where collective control is relevant and feasible. Our goal is not to prescribe specific implementations, but to clarify how different forms of collective control can be described, formalised and evaluated.
Such a formal approach is valuable precisely to pinpoint potential shortcomings in collective control mechanisms and enable a structured and systematic comparison between alternative processes.

We also do not propose or evaluate concrete governance mechanisms, participatory procedures or institutional designs, and also do not present empirical studies of stakeholder involvement. Moreover, while we engage with research on AI alignment, we neither introduce novel alignment methods nor provide a comprehensive review of the alignment literature; rather, we treat alignment as one component within a broader framework of mechanisms for collective control. Finally, our use of social choice theory is intended as a normative and analytical lens, not as a reduction of democratic governance to mere preference aggregation. We view our article as complementary to institutional, empirical and participatory approaches to democratic governance of AI systems.
Lastly, note that we focus on the \emph{institutional} development and use of AI systems, rather than what individual hobbyists, tinkerers and researchers may build in their spare time.

\smallskip

In summary, our contribution thus complements empirical, participatory and regulatory approaches by clarifying where collective decisions arise in the ML pipeline (\Cref{sec:mlpipeline} identifies concrete technical decision points that warrant collective input) and how they can be formalised and evaluated (\Cref{sec:ai-as-scfs} and \Cref{sec:approaches} introduce a common analytical language and normative benchmarks grounded in social choice theory). We believe that regulatory approaches, in particular, would benefit from clearer mappings between high-level governance principles (e.\,g.\ fairness, accountability, oversight) and specific, formally analysable design choices within AI systems. Our framework can help operationalise such principles and may give rise to formally grounded audit metrics that integrate into existing regulatory and accountability frameworks. Our approach is informed especially by prior work in participatory AI (e.\,g.~\cite{sloane22participation}), which emphasises that power and influence are distributed across multiple stages of the ML development pipeline. We build on this insight by systematically identifying these stages and proposing a structured method for analysing and evaluating collective input at each of them.

\section{Collective Decision Points in ML Development}
\label{sec:mlpipeline}

At first glance, AI systems appear more amenable to collective input and decision-making than `classical' algorithmic design, as more people and their data influence system behaviour. However, at present, many consequential development and design choices are still made centrally by a small number of decision-makers~\cite{chaffer2025decentralizedgovernance}. This concentration of power stands in stark contrast to the broad societal impact of these systems. We argue for rethinking how collective input and control can be incorporated into ML development to enable effective collective control of ML.

\begin{figure*}[t]
    \centering
    \includegraphics[width=0.9\linewidth]{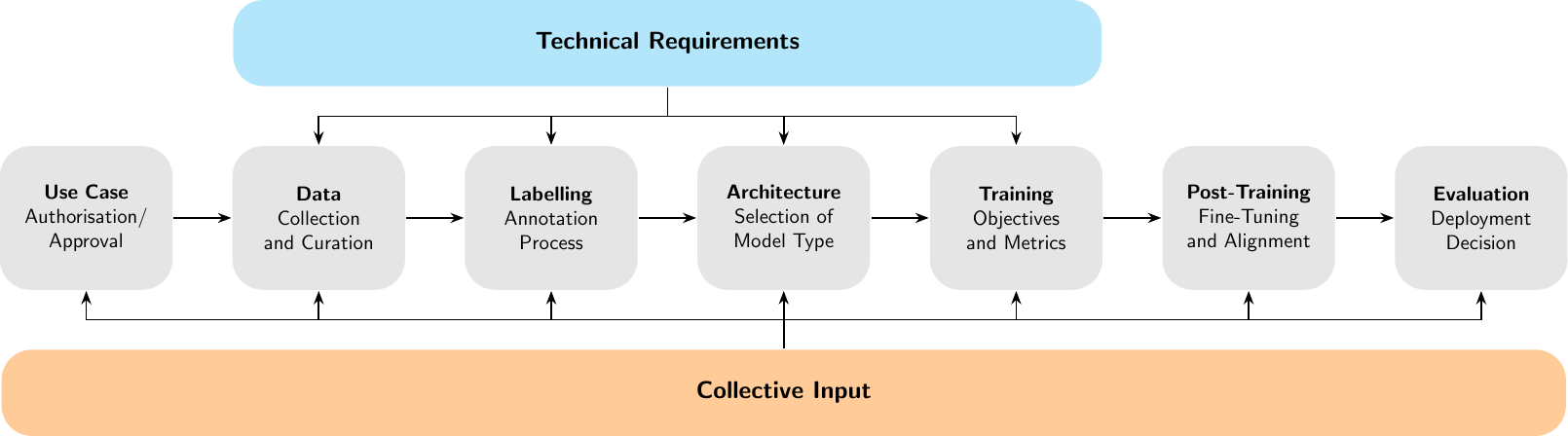}
    \caption{A sketched ML development process; collective input can be meaningfully incorporated at each stage.}
    \label{fig:ml_pipeline}
\end{figure*}

In this section, we will identify various points within the ML development process at which collective input is relevant -- both because stakeholders hold preferences over decisions made at this point and because these preferences actually can be incorporated by collectively controlling such decisions.
To this end, we present a simplified overview of a generic \emph{ML development process} (see \Cref{fig:ml_pipeline}), loosely based on the ML workflow described by Reddi~\cite[Chapter~5]{mlbook2025} and~\cite{AmershiBBDGKNN019}.
Note that -- depending on the concrete case -- some of the development steps may depend on other steps in an iterative process, e.\,g.\ data curation, labelling and architecture selection. However, the mathematical structures that guide development -- the objects of our proposed collective decision-making mechanisms -- remain the same in the presence of such interdependencies.
For clarity, we therefore do not include this additional complexity in the figure.
Conversely, not every development step is relevant for every ML system.

In addition to the development process, we consider the \emph{system behaviour at runtime} and highlight ways of adjusting it in line with stakeholders' preferences.
This section serves three purposes:
First, it illustrates the many opportunities for incorporating collective input across the ML development pipeline; we provide concrete ideas on the `how' in \Cref{sec:approaches}. Second, it demonstrates the complexity and breadth of stakeholders' preferences. Third, it provides background on a generic ML development process, which we use as a scheme for thinking about collective control in ML development in the following.

Beyond the presented schematic overview of the ML development process, we deliberately avoid a formal definition of what we mean by an `AI system' and instead use the term broadly.
We acknowledge that the `black-box' nature of many ML algorithms and AI systems can be a cause for concern, yet we believe that the challenge of collective control arises independently, e.\ g., from the distinction between open-weight and proprietary models.
Before describing the ML development process, we stress that, when we use the term `stakeholders', we deliberately include \emph{all} affected actors, not merely, for instance, economically interested parties. In our usage, `stakeholders' can therefore (and often will) include the public at large, somewhat in contrast to how the term is used in other contexts.

We acknowledge that identifying and empowering structurally marginalised stakeholders requires complementary institutional and participatory work; our formal framework can help evaluate such efforts but cannot replace them. In practice, this broad conception -- necessary to ensure genuinely broad collective control -- typically calls for deliberative elements in preference formation and elicitation, as many affected actors are unlikely to be ML experts. Alternatively, the broader public may be represented in these collective decisions by intermediaries -- such as advocacy organisations, civil society organisations or regulatory bodies -- with sufficient technical expertise. The fact that AI production and deployment are often realised through a global, decentralised supply chain adds additional complexity to identifying and involving affected stakeholders; however, our conceptual suggestions are not dependent on production or deployment occurring within any particular boundary (geographic or otherwise).

We close with the remark that, across the different collective decision points, the preferences of different stakeholders may warrant different relative importance. Accordingly, at each stage, the preference weights assigned to stakeholder groups may need to be adjusted. To prevent collective input from being overridden by technical arguments or other unilateral justifications, these meta-governance questions should themselves be answered democratically, for instance through a deliberative council advised by experts.
While such questions are important for a fully democratic development process, we predominantly focus on the technical problem of aggregating collective preferences at each decision point.

\subsection*{A Schematic ML Development Process}

We now proceed to discuss the ML development process of \Cref{fig:ml_pipeline}, identifying the potential for collective decisions and input in each stage.

\textbf{Authorisation for the use case.}
The first collective decision point concerns whether an AI system should be developed and deployed in a given setting in the first place.
In some contexts, stakeholders may deem any use of AI systems harmful or unethical (e.\,g.\ digital phrenology~\cite{Andrews2024, Stark2022}).
Special attention is required with regards to negative externalities~\cite{hagendorff21}, which are often disregarded or viewed as ``outside'' issues.
The final outcome may be to realise that ``the best solution to a problem may not involve technology'' (cf.\ the Solutionism Trap~\cite{SelbstBFVV19}).

\textbf{Data collection and curation.}
Machine learning models learn from training data and are evaluated on validation and test data. The composition of the training data shapes the resulting system's strengths and weaknesses~\cite{MOHAMMED2025dataquality} and collecting data often entails a variety of societal issues~\cite{paullada2021data}. While the training data itself often comes from a multitude of sources, the mechanisms for data collection and selection are typically devised unilaterally by a small set of decision-makers. We argue that collective input should inform how data is collected and curated as well as the mechanisms that determine which data is ultimately used for training and testing.

\textbf{Labelling and annotation process.}
Many AI systems require training data annotated with labels~\cite{platanios2020learningimperfectannotations}. The complexity of these labels varies substantially across systems, from binary classification to image bounding boxes. Labels are typically provided by annotators who, depending on the task, may be laypeople or experts, and are often distributed, for example, via crowdworker platforms. This process can yield conflicting annotations~\cite{aroyo2015truth,Plank22,platanios2020learningimperfectannotations}, making it necessary to aggregate disagreeing labels fairly and efficiently.
In addition, stakeholders, who are typically not annotators themselves, may have views on the labels and the labelling process: which labels should be used, where and how labels are collected, how trustworthy the respective sources are, and so on. These stakeholder perspectives must be integrated into the procedure to obtain a legitimate labelled dataset.

\textbf{Selection of model architecture.}
The choice of model architecture constitutes a key decision point in the ML development process, as it fundamentally impacts system behaviour as well as system properties such as explainability, bias, accessibility and environmental impact~\cite{INDYKOV2025architecturaltactics,delrey2025estimatingdeeplearningenergy}.
While the technical details of architectures are typically too specialised for direct public decision-making, the public may nonetheless have preferences over system properties that are affected by this choice.
Moreover, specialised interest groups representing the public may hold specific views on architectural decisions. Consequently, this stage admits collective input in various forms.

\textbf{Training objectives and metrics.}
Similar to the model architecture, collective decisions are equally relevant for the choice of loss/reward functions, training objectives, target accuracy and related metrics. These choices operationalise what the system is optimised to do and, as a result, can embed normative trade-offs between competing goals~\cite{Ji2025alignment, CorbettDavies2023} (e.\,g.\ accuracy, safety, fairness and cost). Further, understanding human psychology in how goals are formed and pursued is essential for co-selecting objective functions for AI systems~\cite{Gabriel2020}.
Research is ongoing on bridging the gap between human and artificial objective functions~\cite{gershman2025subjectivefunctions}.

\textbf{Fine-tuning and alignment.}
In the ML industry, it is common to pre-train models, and especially LLMs, on a corpus of data and then fine-tune them (for instance, using reinforcement learning) to improve their usefulness and align them with human values and ethical principles~\cite[Chapter~17]{mlbook2025}. Some approaches derive these values from stakeholders, employing simple ad-hoc methods to elicit and aggregate preferences without axiomatic considerations~\cite{Anthropic2024}. Since questions of AI ethics and value alignment are inherently subjective, fair collective decision-making mechanisms are crucial for determining the values a system should be aligned with.

\textbf{Evaluation and deployment decision.}
There is a need to establish a collective consensus -- at the beginning of the development process -- on the metrics and thresholds that justify deploying the system. However, even once development is complete and the system is tested against these thresholds, it may not be clear-cut whether a given threshold has been met; moreover, the initial decision may have left some leeway. In either case, it may be necessary to make an explicit collective decision on whether the requirements have, in fact, been satisfied.

\textbf{Refinement of model output/behaviour at runtime.}
We turn our attention to the deployed system. In deployment, AI systems are often embedded in larger systems that further refine their behaviour, enable them to incorporate real-time information or adjust their outputs to a specific deployment context. One widely used approach is retrieval-augmented generation (RAG), in which a generative model retrieves external information at runtime, allowing it to integrate up-to-date data when generating outputs. Decisions made at this stage concern the composition of the data sources used as runtime input after deployment. They also concern, similarly to the fine-tuning stage, the ethical principles used to guide or constrain behaviour: Which responses to a given input best align with stakeholders' ideals and how should conflicts between principles be handled at runtime?

Finally, we would like to mention monitoring and maintenance of a deployed system. After deployment, ML models are susceptible to performance degradation due to concept or data drift~\cite{Lu2019}. Furthermore, AI audits -- external or internal evaluations for bias, fairness and compliance -- are becoming an important mechanism for post-deployment accountability~\cite{Raji2020,CostanzaChock2022}. As these challenges are primarily related to macro-level governance and less amenable to the type of collective decisions discussed here, we consider them beyond the scope of this article.

\section{End-To-End Collective Control: AI Systems as Social Choice Functions}\label{sec:ai-as-scfs}

In the following, we use the lens of social choice theory to develop a simple mathematical model for collective control of AI.
To this end, we view AI systems as social choice functions. This perspective makes the collective decision-making dimension of a model's end-to-end behaviour explicit,  allows us to reason about it formally, and gives rise to various evaluation criteria.
In this section, we treat the AI system as a black box and focus purely on its end-to-end behaviour; in this sense, our model can be understood as a model for \emph{collective control at deployment}. In contrast, the next section focuses on collective control across the various stages of ML development.

\subsection{The Formal Model}
\begin{figure*}[t]
    \centering
    \includegraphics[width=0.8\linewidth]{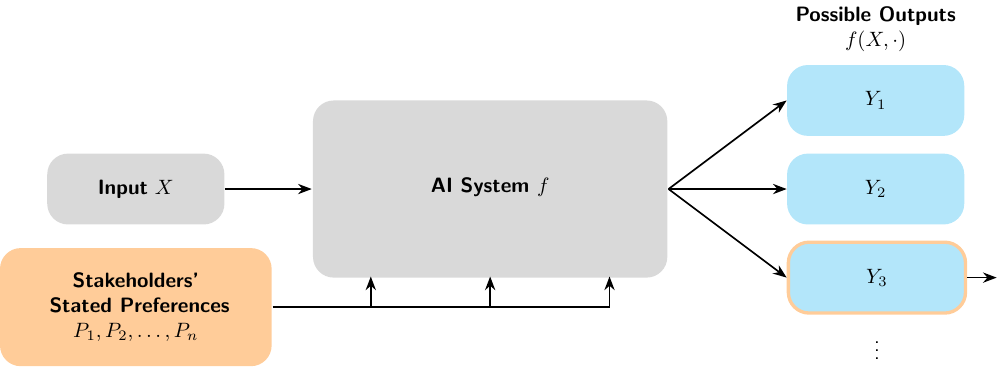}
    \caption{A sketch of our mathematical model, where we view an AI system $f$ as a social choice function.}
    \label{fig:model}
\end{figure*}

Let $f$ be the AI system under consideration. Presented with some input $X$, system $f$ will produce some outcome $Y$. We assume that there are $n$ stakeholders, each of which provides their own input $P_i$, $i\in \{1,\dots,n\}$, to $f$. This input of stakeholders can be viewed as their \emph{stated preferences} about the output of $f$. These stated preferences may or may not depend on $X$. Joining these components, we obtain $f(X,P_1,\dots,P_n)=Y$.
If we want to refer to the range of possible outcomes for an input $X$, we write $f(X,\cdot)$.
The outline of this model is sketched in \Cref{fig:model}.

We do not impose a specific structure or model on stated preferences. However, we assume that we can use $P_i$ to compare two possible outcomes $Y_1,Y_2$ on some input $X$ and write $Y_1\succeq^X_i Y_2$ if $P_i$ induces that outcome $Y_1$ is weakly preferable to outcome $Y_2$ on input $X$ by stakeholder~$i$.
 We write $Y_1\succ^X_i Y_2$ to denote that $Y_1$ is strictly preferred by stakeholder~$i$.\footnote{\ Under realistic assumptions, it will require additional conditions and mechanisms to use $P_i$ for pairwise comparisons ($Y_1\succeq^X_i Y_2$). We assume that $P_i$ is a much simpler object than a full preference relation, e.\,g.\ a set of labelled comparisons or a natural language description, and that pairwise judgments are derived via ML alignment techniques such as reward modelling or constitutional approaches; for a comprehensive survey, see~\cite{Ji2025alignment}. If an evaluation function or reward model maps possible outcomes to real numbers (their utility for stakeholder~$i$), the corresponding utility function naturally implies a weak ordering of outcomes and thus fits our model.}

\begin{example}\label{ex:student-chatbot}
    Consider an LLM-based chatbot to help secondary school students identify university education programmes; this is our AI system $f$. Several stakeholder groups provide stated preferences:
    \begin{enumerate}
        \item $P_1=\texttt{"Provide information about public and private}$ $\texttt{institutions."}$ (educational authorities)
        \item $P_2=\texttt{"Include scholarship opportunities and admission requirements."}$ (parents' associations)
        \item $P_3, \dots, P_{10}$ are educational institutions, each with a preference for its own agenda and a preference for having this agenda prominently reflected in the response.
    \end{enumerate}
    Given input $X=\texttt{"What can I study if I like playing and}$ $\texttt{thinking about video games?"}$, the system may respond with $Y=f(X,P_1,\dots,P_{10})=\texttt{"Here are the top three CS degrees}$ $\texttt{in the country: ..."}$. This output can now be evaluated (with appropriate methods) with respect to how well it respects stakeholder preferences.
\end{example}

\noindent The following assumptions underline our model:
\begin{enumerate}
    \item We will not consider the entity that is actually in control of (or developing) the AI system $f$, typically a company, as part of the set of stakeholders. They have a special role as they ultimately decide on what exactly the function $f$ should look like and thus very different possible actions and points of influence than other stakeholders.
    \item We assume that the ranges of values for both $X$ and $Y$ are very large; many inputs and outcomes are possible.
    \item Due to the probabilistic nature of most AI systems, $f(X,\cdot)$ might be large even if $f$ does not allow for stakeholder input (or simply ignores it).
    \item The task of an AI system $f$ typically implies hard constraints on its output, which stakeholders may not be fully aware of. In other words, given some input $X$, we do not assume that even a single stakeholder's preference is realisable, i.\,e.\ we might be unable to select the most preferred responses of a stakeholder on input $X$.\footnote{\ In the language of social choice theory, this is a violation of non-imposition: Not all alternatives can be chosen.}
    \item The model is agnostic as to the technical details of how stakeholder preferences are incorporated.
    \item In this section, we assume that stakeholders' preferences are purely over the outcomes returned by the model. This is in contrast to \Cref{sec:approaches}, where we discuss more fine-grained opportunities for collective control.
\end{enumerate}

\noindent Let us contrast this simplistic model of an AI system $f$ with the definition of a \emph{social choice function (SCF)}~\cite{Moul88a,brandt2016handbook}. An SCF gets as input a set of candidates $C$ and a set of voters $1,\dots,n$, each expressing their preferences on this set of candidates in some mathematical structure, e.\,g.\ a linear order. The SCF then aggregates the voters' preferences by selecting one candidate from $C$, i.\,e.\ it chooses an optimal candidate from $C$ according to voters' preferences.
Clearly, not all functions of this form are actually sensible SCFs. For example, an SCF could always return the most preferred alternative of voter~1, which is known as a \emph{dictatorial} social choice function.

Importantly, fixing the input $X$, we can view $f$ as a social choice function, as it selects an outcome $Y$ from the set of all possible outcomes, while having access to the stakeholders' preferences over such outcomes; stakeholders in our model can be viewed as voters and outcomes as candidates.
However, currently, AI systems are oftentimes not designed to be a social choice function: They completely ignore democratic input or fundamentally lack the possibility of providing such input in the first place --
a behaviour that may be seen as an even more extreme form of dictatorship.

\subsection{The Axiomatic Perspective from Social Choice Theory}\label{sec:axioms}
In social choice theory, the established method to distinguish and analyse SCFs is axiomatic analysis. This means that \emph{axioms}, i.\,e.\ desirable mathematical properties of SCFs, are proposed and then used to classify and judge SCFs.
The above-described modelling of the end-to-end behaviour of AI systems as an SCF allows the translation of social choice axioms to check the behaviour of AI systems.
We believe that such an axiomatic approach could be highly valuable in evaluating how AI systems take stakeholders' preferences into account and implement collective control. In the following, we discuss major social choice axioms and their meaning in the context of controlling AI systems.\footnote{\ There is a multitude of further axioms, pertaining, e.\,g.\ to monotonicity, consistency or clone-proofness. Their importance is fundamentally dependent on the application context; evaluating their relevance for the collective control of AI context is an important direction for future work.}
We first present these axioms as binary properties and later explain how they can be turned into quantifiable properties (enabling the measurement of the degrees of fulfilment).

\textbf{Anonymity.}
Anonymity is one of the most basic social choice axioms~\cite{Moul88a}. It states that all voters are treated equally or, framed differently, voters' identities are not taken into account in the role their preferences play for the outcome. This axiom rules out dictatorial SCFs, but also weighted SCFs, where voters have different weights (e.\,g.\ countries according to their population or voters according to their social class). In our context, it is \emph{prima facie} unclear whether an anonymity requirement is actually meaningful. First, stakeholders may be affected to different degrees by an AI system and thus actually deserve different consideration; such an approach would be ruled out by anonymity. Second, even if anonymity holds, it could be guaranteed vacuously by simply ignoring all stakeholders -- if $f(X,P_1,\dots,P_n)=f(X)$, all stakeholders are treated equally (i.\,e.\ ignored). Given these observations, we are rather interested in an axiom that guarantees meaningful impact on $f$ for all stakeholders. The next axiom, participation, is a first step in this direction.

\textbf{Participation.}
Counter-intuitively, for some SCFs, it may be harmful to participate in a collective decision-making process. For instance, it may be the case that stakeholder~1 is less satisfied with the outcome $f(X,P_1,P_2,\dots,P_n)$ than with the outcome $f(X,P_2,\dots,P_n)$ (where stakeholder~1 does not participate).
This so-called ``no-show paradox'' is well-known in social choice theory~\cite{moulin1988condorcet}. The participation axiom is the axiomatic property requiring that this kind of behaviour cannot occur: Participation must never be disadvantageous.
In the scope of AI systems, this property retains its meaning as a fundamental guarantee: If an AI system allows for collective input of preferences, then providing one's preferences must not lead to a worse outcome for participants. Note that this property is far from being trivial; an AI system may use stated preferences as information to advance its core objective, which can contradict individual stakeholders' interests.
In mathematical notation, participation requires that for all stakeholders $i\in\{1,\dots,n\}$,
\begin{align}
    \text{for all stakeholders }i\in\{1,\dots,n\},\ f(X,P_1,\dots,P_{i-1},P_{i+1},\dots,P_n) &\succeq^X_i f(X,P_1,\dots,P_n).
\label{eq:participation}
\end{align}

\textbf{Unanimity and the majority criterion.}
In many settings, in case a (large) majority of voters are in agreement with each other, they should be able to decide the outcome of an SCF. This idea is captured by several axioms in social choice theory. Most important, and most uncontroversial, is the principle of unanimity: If all voters unanimously agree on a candidate (as their most preferred one), an SCF should select it. The majority criterion is a stronger version: If a (strict) majority of voters agree on a candidate, an SCF should select it.

Adapting these notions to our context, we first note that it is implausible to assume that stakeholders provide sufficiently detailed preferences to even identify a single most-preferred outcome $Y$ for each stakeholder. However, they may unanimously prefer an outcome $Y_1$ over an outcome $Y_2$ on some input $X$. If $Y_1$ is valid (i.\,e.\ contained in $f(X,\cdot)$), $f$ should not select $Y_2$.
Accordingly, we can formalise unanimity as: If $Y_1 \succ^X_i Y_2$ for all $i\in\{1,\dots, n\}$ and $Y_1\in f(X,\cdot)$, then $f(X,P_1,\dots,P_n)\neq Y_2$.
The majority criterion can be defined analogously by requiring $Y_1 \succ^X_i Y_2$ to hold only for a majority of stakeholders (and not all of them).

\textbf{Pareto efficiency.}
This basic axiom pertains to efficiency, i.\,e.\ an SCF's ability to identify ``better'' outcomes. Pareto efficiency (phrased in our model) requires that if for two possible outcomes $Y_1,Y_2 \in f(X,\cdot)$
everyone weakly prefers $Y_1$ over $Y_2$ and at least one stakeholder strictly prefers $Y_1$ over $Y_2$, then $Y_2$ should not be selected. Formally,
\begin{equation}\label{eq:pareto}
    \text{if }Y_1 \succeq^X_i Y_2 \text{ for all }i\in\{1,\dots,n\} \text{ and }Y_1 \succ^X_i Y_2 \text{ for some }i\in\{1,\dots,n\}, \text{ then } f(X,P_1,\dots,P_n)\neq Y_2\text{.}
\end{equation}

Pareto efficiency can be regarded as a stronger version of unanimity as defined above. Note that this axiom is not about fairness; an outcome that is ideal for one stakeholder but terrible for everyone else may be Pareto-efficient.

\subsection{From Axioms to Evaluation Criteria}
As argued above, axiomatic properties of SCFs are also applicable to the evaluation of AI systems. However, these axioms are framed as binary properties -- either satisfied or not. As ML-based systems are typically non-deterministic and highly complex, it is unreasonable to demand or formally establish perfect guarantees. Indeed, also for (much simpler and deterministic) well-established SCFs, it has been observed that some basic properties only hold ``most of the time''~\cite{regenwetter2006behavioral,friedgut2011quantitative, gehrlein2010voting}.
However, it is easily possible to turn binary axioms into quantitative measures, as we can measure (count) how often a given axiomatic property is violated by an AI system.
For example, let us consider participation (Property~\eqref{eq:participation}). If we want to test the degree to which an AI system~$f$ satisfies participation, we can test it on a set $\mathcal{I}$ of inputs of the form $(X,P_1,\dots,P_n)$ and
record for how many pairs $(X,P_1,\dots,P_n)\in\mathcal{I}$ and $i\in\{1,\dots,n\}$ Property~\eqref{eq:participation} is satisfied, resulting in a quantitative participation measure that can be used as an evaluation criterion.\footnote{\ Such a measure would be highly dataset-dependent, necessitating the development of meaningful benchmarks for the impact of collective input.}

For Pareto efficiency, unanimity and majority, we would need to follow a slightly different approach. These axioms reason over any pair of two outcomes fulfilling certain preference constraints. However, iterating over all pairs of outcomes is infeasible for most AI systems.
We propose two alternative approaches:
\begin{enumerate}
    \item Set $Y_2=f(X,P_1,\dots,P_n)$ and sample $Y_1$ from a suitable distribution over $f(X,\cdot)$. Then count how often the pair $(Y_1,Y_2)$ witnesses a violation of the respective axiom. Note that this approach requires a very good understanding of the output space, as otherwise a suitable $Y_1$ may not be sampled (even if suitable choices for $Y_1$ are relatively common). Alternatively, stakeholders themselves may suggest alternative outcomes -- or even sampling procedures.
    \item A more practical approach is to instead set $Y_2=f(X)$, i.\,e.\, the outcome obtained when ignoring stakeholder input. If this outcome is socially preferable to $Y_1=f(X,P_1,\dots,P_n)$ under the respective axiom, there is indeed something amiss -- the collective input has led to an even worse outcome.
\end{enumerate}

\subsection{Temporal Fairness and Group Fairness}
Finally, let us discuss the issue of fairness, which becomes especially relevant when considering the behaviour of AI systems over longer periods of time.
As it is often impossible to satisfy all stated preferences, we can ask: Are there (groups of) stakeholders who are systematically disregarded over a longer time frame?
Note that to incorporate a temporal dimension in our model, all parameters can be taken to be sequences: $X$ can be a sequence of inputs, $P_1,\dots,P_n$ can be sequences of stated preferences (as they may change over time) and $Y$ is a corresponding sequence of outputs.
The problem of an AI system given a sequence of $(X,P_1,\dots,P_n,Y)$-tuples is closely related to a recent research direction called \emph{temporal fairness}~\cite{aaai/perpetual,aaai/proportional-perpetual-voting,parkes2013dynamicscevolving,Elkind2025temporalproportionality,elkind2025temporalchores,freeman2017fairscdynamic}. The goal of temporal fairness is to fairly balance interests over a sequence of decisions. This would be highly desirable for fairness in AI systems. Ideally, such a mechanism would also take intensities into account, e.\,g.\ to prioritise stakeholders who are especially affected by a certain outcome.

\begin{example}\label{ex:tf}
    Returning to \Cref{ex:student-chatbot}, note that the given query~$X$ may not be relevant to all education institutions -- some may not offer relevant programmes. If, however, a university is repeatedly not listed despite offering suitable programmes relevant to the prompt, this constitutes a violation of some definition of temporal fairness.
\end{example}

\noindent \Cref{ex:tf} would violate a notion of \emph{individual fairness}. Even more attention is typically paid to \emph{group fairness}. For example, there might be a systematic bias towards larger educational institutions, neglecting smaller, more specialised universities. Group fairness is most often studied through the lens of proportionality: If a group corresponds to $x\%$ of the stakeholders (appropriately weighted), it should be able to influence at least $x\%$ of all decisions. The exact formulation of proportionality depends on both the form of preferences used (0/1 approval preferences, rankings, utility functions, etc.), the social choice mechanism in use (single-winner, multi-winner, clustering, etc.), as well as the specifics of the settings (sequential decisions, parallel decisions, fixed or variable sequence lengths, etc.). The surveys by \citet{BoehmerN21} and \citet{ElkindOT24} discuss recent advances in this direction.

\section{Collective Control in the ML Development Process}\label{sec:approaches}

Complementing our framework for evaluating collective control of the end-to-end behaviour of AI systems at deployment, we now propose avenues for collective control throughout the ML development process, following the model described in \Cref{sec:mlpipeline}.
We will also discuss collective decision points in deployment here, but (unlike our model in the previous section) go beyond reasoning solely about the ultimate outputs of deployed models.
This section shows how social choice can provide a new perspective on collective control in ML development by connecting the decision points identified in \Cref{sec:mlpipeline} with well-studied collective decision-making problems.
Specifically, we consider the authorisation for the use case (\Cref{subsec:Authorisation}), data collection and curation (\Cref{subsec:data-collection}), the labelling and annotation process (\Cref{subsec:labeling}), the selection of model architecture and training parameters (\Cref{subsec:architecture}), fine-tuning and alignment (\Cref{subsec:alignment}) and refinement of model behaviour at runtime (\Cref{subsec:runtime}).

For each decision point, we specify the key entities for collective input via social choice: the set of available \emph{alternatives}, the \emph{kind of preferences} that stakeholders may have and the \emph{form of outcome}.
Although we focus on the question of how to aggregate elicited preferences, we emphasise that some decision mechanisms will need to be embedded in structures that enable stakeholders to develop meaningful and informed preferences. These might include deliberative councils or other processes that allow stakeholders to acquire relevant domain knowledge and engage in debate.

\subsection{Authorisation for the Use Case}\label{subsec:Authorisation}
Setting boundaries for AI usage and applications is an integral first step of collectively controlling AI systems and their integration into society.
Such considerations are complementary to the framework in \Cref{fig:model}, as authorisation decisions involve socio-technical and environmental questions that go beyond the behaviour of the AI system itself.

Collective control of authorisation decisions can take various forms: from a simple yes/no vote before development begins to more complex authorisation decisions that are contingent on the behaviour of the developed system.
We focus on the latter, which requires more intricate procedures for collective control.
Specifically, stakeholders' preferences here can be modelled as a set of minimal criteria or requirements that must be met for them to approve an AI system. For instance, for a job-application filtering system, stakeholders may only consent to the deployment of a system if it can be shown to exhibit sufficiently low bias. Since this often cannot be guaranteed \emph{a priori}, the goal at this initial decision point is to collectively establish metrics and binding target thresholds that the finished system must satisfy to be deployed.
This, in turn, requires aggregating potentially conflicting stakeholder views about what constitutes acceptable minimum thresholds.
Selecting authorisation requirements is related to several collective decision problems:
\begin{itemize}
    \item Selecting a (small) set of target metrics: This corresponds to multi-winner voting~\cite{Lackner22abcvoting, faliszewski2017multiwinner} (a well-studied setting in social choice in which a fixed-size subset must be selected from a set of candidates) and allows for fair representation of affected groups.
    \item Selecting numerical thresholds: Here, median-based procedures can be used to limit the influence of outliers and strategic voting~\cite{moulin1980strategy,FREEMAN2021105234}.
    \item Combined approaches that select metrics and thresholds simultaneously, ideally with fairness guarantees: This is an open problem; we are not aware of prior work studying such combined settings.
\end{itemize}

\noindent As AI systems are commonly repurposed, effective collective control in fact requires separate authorisation for each use case, with stakeholders' preferences potentially varying substantially across use cases.
We consider this differentiation highly relevant; for now, however, we assume that each individual application of a general-purpose system requires its own completely independent authorisation, each constituting a separate collective decision-making problem.

\subsection{Data Collection and Curation}\label{subsec:data-collection}
The central questions around data collection concern the selection of datasets for training and validation of AI systems: What are fair and efficient procedures that allow stakeholders to collectively decide which data to use? How should harmful data be excluded?
Accordingly, the collective decisions made at this stage have implications ranging from the behaviour of the model trained on the selected data
to more `procedural' aspects such as permissible modes and sites of data collection and imposed standards of data protection.
Collective control over data collection should therefore go beyond approaches such as the European Union's GDPR: It would mean that those affected by data collection (which would be included in our interpretation of stakeholders) not only decide individually whether to provide their own data, but also collectively which \emph{kinds} of data may be used for a given AI system or use case. This would give them the ability to weigh potential benefits against collective sacrifices in privacy and other ethical dimensions.

Note that stakeholders are unlikely to have preferences over individual data points, but rather over the overall composition of the dataset, or, more abstractly, over properties of the dataset
(such as proportional representation or low bias).
Collective decisions about dataset composition can be captured by social decision schemes (SDSs) based on portioning with cardinal preferences~\cite{elkind2024portioningcardinal}. In our setting, this would require us to partition data into types (e.\,g.\ based on sources, media type or informational content) and eliciting stakeholders' preferences over the desired type composition of the dataset. An SDS then produces a distribution over data types that is, in some sense, optimal (e.\,g.\ proportionally representative or maximising a notion of social welfare). This distribution can be interpreted either as the target composition of the dataset or as weights assigned to the different data types.

A further aspect of collective control concerns excluding undesirable (i.\,e.\ discriminatory or otherwise harmful) data. Again, stakeholders may disagree about what counts as harmful and about which thresholds should trigger exclusion. Here, we can model stakeholders as having dichotomous preferences, partitioning data into desirable and undesirable. One way to aggregate such dichotomous preferences is via portioning functions based on approval ballots. Rather than submitting individual data points, stakeholders could again be asked to specify \emph{types} of data they do not want included in the dataset~\cite{bogomolnaia2004dichotomousSDS}. This could be integrated with the portioning approach above into a joint pre-selection step.

In the context of our model from \Cref{fig:model}, collective decisions around data collection affect the range of possible outputs of the AI system and can therefore additionally contribute to the alignment of the output with stakeholders' preferences. From an axiomatic perspective, diversity and proportional representation constraints can ensure a degree of representation for minorities in the data~\cite{aaai/BFILS-diversity}.

\subsection{Labelling and Annotation Process}\label{subsec:labeling}

As data labels and annotations are often obtained in a decentralised fashion using multiple annotators, there may be disagreement about the appropriate label for a data point. This motivates either (i) learning methods that can directly accommodate disagreeing information~\cite{uma2022disagreementlearning} or (ii) an aggregation step that resolves disagreement into a single training target~\cite{platanios2020learningimperfectannotations,braylan2021aggregate_merge_match,yin2017aggregatingcrowdwisdoms}. Two common approaches for (ii) are to form either \emph{hard} or \emph{soft} labels.

The more common approach produces a single, hard label per data point. The resulting task of inferring a ``true'' label from a set of annotations has already been approached through a social choice lens: Annotators act as voters who report their perceived best-fitting label from a fixed label set. A standard aggregation procedure is (weighted) plurality or majority voting, possibly reweighting annotators by estimated reliability~\cite{aydin2014crowdsourcing,Burke2020bayesianinference}. However, also beyond the currently used plurality voting, social choice theory offers a large variety of voting rules that could be applied, with different informational requirements on annotators~\cite{brandt2016handbook}. This perspective is closely related to \emph{epistemic voting}, which also explicitly assumes an underlying ground truth and designs voting procedures that recover that ground truth, where the optimal epistemic voting rule depends on the assumed noise model of the annotators' reports~\cite{Pivato2013comsoc_mle,allouche2022multiwinnerapprovalvotinggoes}.

In contrast, soft-label approaches avoid committing to a single label and instead represent the target as e.\,g.\ a distribution over labels. This is particularly natural in settings where the ``correct'' label is subjective and shaped by annotators' or users' experiences~\cite{lakoff1987categoryperception}. Soft labels can therefore preserve valid disagreement and have also been observed to improve predictive performance in several settings~\cite{Uma2020,Mandal2024,Peterson2019}. Soft labels may be obtained implicitly by aggregating hard annotations into empirical label distributions or explicitly by eliciting probabilistic reports or richer preference information from annotators; the fair and efficient aggregation of such information is an active research topic~\cite{yin2017aggregatingcrowdwisdoms,braylan2021aggregate_merge_match,Burke2020bayesianinference}. Social choice theory can again contribute to this problem: A broad range of multi-winner and probabilistic voting rules can be used to aggregate reports into a set of winning labels or a full distribution.

Modelling annotation aggregation as a social choice problem highlights the subjectivity of human perception and categorisation, rather than treating annotators purely as noisy sensors of an objective truth. This framing is especially appropriate in applications where preferences are inherently subjective and where disagreement can be legitimate. At the same time, deploying voting rules beyond plurality often requires eliciting richer preference information; there is ongoing work on procedures that remain compatible with highly incomplete or otherwise coarse preference data~\cite{halpern2023representation,lu20elicitation_multiwinner}.

\subsection{Selection of Model Architecture and Training Parameters}\label{subsec:architecture}
When selecting a model architecture as well as the specific objectives, parameters and metrics used to train and evaluate the system, considerations such as resource consumption or explainability often need to be balanced against accuracy and overall system capabilities~\cite{delrey2025estimatingdeeplearningenergy,INDYKOV2025architecturaltactics,strubell-etal-2019-energy}. This includes highly technical questions about the hyperparameters used for training, such as training data batch size or the choice of optimiser.
Because choices of architecture and hyperparameters must match the task at hand and satisfy technical feasibility constraints, developers can easily appeal to these requirements to disregard public opinion altogether. However, the consequences of architecture choices (e.\,g.\ environmental consequences) can very well be subjected to public debate~\cite{rehak25}.

This problem aligns with the standard single-winner voting setting in social choice theory (one of many feasible architectures, optimisers or parameter values needs to be selected) and thus calls for a social choice function to aggregate stakeholders' preferences into a collective decision.
For this process, stakeholder preferences may be elicited in various ways, for example as minimum requirements regarding resource efficiency or system capabilities.
A key challenge in the design of a targeted social choice function is how to balance accuracy and capabilities against stakeholders' preferences. One approach is to restrict the set of eligible architectures (i.\,e.\ the candidates presented to the social choice function) to those that are powerful enough to solve the task. Another is to use stakeholders' preferences primarily to rule out socially undesirable architectures (for instance, via concepts such as the proportional veto core~\cite{moulin1981proportional}) rather than granting them full control over the final choice.

\subsection{Fine-Tuning and Alignment}\label{subsec:alignment}
A major challenge for collective input at this stage is aggregating diverse human values and ethical principles. Notably, fine-tuning and alignment are already typically based on some form of human preference data.
However, existing approaches often neglect the collective and pluralistic nature of such data: They ignore questions of representativeness and treat disagreement as noise rather than as valid, conflicting views that necessitate mediation~\cite{sorensen2024pluralisticalignment}.
We review alignment approaches in ML research and discuss how they could be improved using aggregation methods from social choice theory. Since stakeholders' preferences at this stage typically induce preferences over system outputs, this step can also be understood through our model in \Cref{fig:model}.

\emph{Reinforcement learning from human feedback (RLHF)} is widely used to align LLMs. Here, human feedback typically takes the form of pairwise preferences over model responses, collected from paid annotators, and disagreement between annotators is resolved only implicitly.
Although eliciting and aggregating preferences over different alternatives lies at the heart of social choice research, social choice mechanisms are rarely used in the context of RLHF. The use of such mechanisms to evaluate or improve the aggregation of human feedback has been proposed by Conitzer et al.~\cite{Conitzer2024} under the name \emph{reinforcement learning from collective human feedback (RLCHF)}.

\emph{Constitutional AI} uses a set of values -- a ``constitution'' -- to represent preferences about LLM behaviour~\cite{Bai2022}. The model is then aligned to this constitution using a reinforcement learning approach. Typically, the constitutional AI literature assumes that the constitution is given, which in practice might often amount to a company board deliberating and deciding on it. Notably, \citet{Anthropic2024} propose a process for eliciting the contents of the constitution from collective input under the name \emph{collective constitutional AI (CCAI)}, but this process provides no formal guarantees and does not draw on established results from social choice.
The problem of selecting a constitution based on stakeholders' preferences can be viewed as a multi-winner voting problem; in our setting, the candidates are values and the selected subset is the constitution. We view the mechanisms and axioms developed for multi-winner voting as a powerful toolbox for evaluating and potentially designing mechanisms for the collective control of constitutions.

\subsection{Refinement of Model Behaviour at Runtime}\label{subsec:runtime}
At runtime, a model's behaviour may be refined, for instance, by incorporating additional real-time information. However, executing complex collective decision procedures at runtime is typically infeasible. Collective input must therefore be aggregated into more abstract principles that can be imposed at runtime.
For example, with respect to real-time information sources, stakeholders may have preferences over the desired composition, analogous to decisions about training data, e.\,g.\ with respect to data formats, types or sources, and may require diversity across geography or political bias. As with training data selection, we can use cardinal preference and veto- or approval-based portioning methods to execute and evaluate these collective decisions.

Runtime refinement can also be used to better align an AI system's outputs with fairness or representation concerns or with collectively established values and principles.
These should again be obtained from collective input using a preference aggregation mechanism, similar to the constitutional values for the CCAI approach.
A version of this idea has been implemented for recommender systems: \emph{Social Choice for Recommendation Under Fairness--Dynamic (SCRUF-D)}~\cite{scruf_d} is a framework that incorporates independent, heterogeneous and dynamic fairness concepts to refine recommender system output using a multi-agent social choice approach. Social choice mechanisms are used both to allocate compatible fairness agents to each recommendation instance and to aggregate the recommender system's ranking with rankings preferred by these fairness agents. In SCRUF-D, the fairness agents are selected by domain experts rather than by general stakeholders, but this could be adapted. For example, instead of representing different fairness notions, each agent could represent a (group of) stakeholder(s) with preferences over the output.

\section{Conclusion}\label{sec:conclusion}
We complement existing approaches to democratic control of AI with a more technical, yet still holistic, view grounded in social choice theory.
On the one hand, we develop a formal model of end-to-end collective control that frames AI systems as social choice functions. This framing allows us to adapt axioms from social choice theory (such as participation, Pareto efficiency and temporal fairness) as evaluation criteria for how well AI systems respect stakeholder preferences. On the other hand, we identify stages of ML development that are particularly salient for incorporating collective input and relate them to established social choice problems such as multi-winner voting and portioning.

Naturally, our approach connects to a range of related work. For instance, the Democracy Levels framework~\cite{Ovadya2025} offers a method for evaluating how democratic an AI ecosystem is, but remains agnostic about which decisions should be made democratically. We focus on precisely this question and argue that, throughout the ML development pipeline, there are decisions in which the public and/or relevant stakeholders have a legitimate interest in influencing the outcome, especially in high-stakes applications. While such a democratised process does not guarantee societally beneficial outcomes, it is a prerequisite for meaningful control of AI.

Several research directions are needed to further develop collective control of AI. First, we have not specified who exactly the stakeholders are; this will clearly depend on the precise setting. Open questions on this issue include how to identify stakeholders at different decision points, how to assess their relative importance and how to design mechanisms that are appropriate for all relevant groups. Second, an important next step is to instantiate the social choice model developed in \Cref{sec:ai-as-scfs} for concrete AI applications and to develop benchmarks that measure axiomatic violations empirically; both steps potentially also require methodological advances in social choice theory to be able to reflect the details and complexity of a given AI system adequately. Third, preference elicitation for non-expert stakeholders poses substantial challenges, particularly for technically complex decisions; methods that enable meaningful participation without requiring deep technical expertise remain to be developed. Finally, it would be interesting to explore how to incorporate axiomatic desiderata, such as participation or temporal fairness, into audit frameworks~\cite{Raji2020,CostanzaChock2022}.

\section*{Generative AI Usage Statement}
The authors confirm that we did not use any generative AI tools during any stages of this research work.

\section*{Acknowledgments}
The first author was funded by the Austrian Science Fund (FWF) through project 10.55776/COE12 ``Bilateral AI''. The third author was funded by the Austrian Research Promotion Agency (FFG) through project FO999904624 ``FAIR-AI''. The financial support by the Austrian Research Promotion Agency and the Austrian Science Fund is gratefully acknowledged. We are grateful to the anonymous referees for suggesting numerous improvements to both the content and the presentation of this paper.

\bibliographystyle{plainnat}
\bibliography{Bibliography}

\begin{thebibliography}{102}
\providecommand{\natexlab}[1]{#1}
\providecommand{\url}[1]{\texttt{#1}}
\expandafter\ifx\csname urlstyle\endcsname\relax
  \providecommand{\doi}[1]{doi: #1}\else
  \providecommand{\doi}{doi: \begingroup \urlstyle{rm}\Url}\fi

\bibitem[Aird et~al.(2025)Aird, Farastu, Sun, Stefancov\'{a}, All, Voida,
  Mattei, and Burke]{scruf_d}
Amanda Aird, Paresha Farastu, Joshua Sun, Elena Stefancov\'{a}, Cassidy All,
  Amy Voida, Nicholas Mattei, and Robin Burke.
\newblock {Dynamic Fairness-Aware Recommendation Through Multi-Agent Social
  Choice}.
\newblock \emph{ACM Trans. Recomm. Syst.}, 3\penalty0 (2), 2025.
\newblock \doi{10.1145/3690653}.

\bibitem[Allouche et~al.(2022)Allouche, Lang, and
  Yger]{allouche2022multiwinnerapprovalvotinggoes}
Tahar Allouche, Jérôme Lang, and Florian Yger.
\newblock {Multi-Winner Approval Voting Goes Epistemic}.
\newblock In \emph{Proceedings of the 38th Conference on Uncertainty in
  Artificial Intelligence}, UAI 2022/Proceedings of Machine Learning Research
  180, pages 75--84, Cambridge, MA, 2022. JMLR and Microtome Publishing.
\newblock URL \url{https://proceedings.mlr.press/v180/allouche22a.html}.

\bibitem[Amershi et~al.(2019)Amershi, Begel, Bird, DeLine, Gall, Kamar,
  Nagappan, Nushi, and Zimmermann]{AmershiBBDGKNN019}
Saleema Amershi, Andrew Begel, Christian Bird, Robert DeLine, Harald Gall, Ece
  Kamar, Nachiappan Nagappan, Besmira Nushi, and Thomas Zimmermann.
\newblock {Software Engineering for Machine Learning: A Case Study}.
\newblock In \emph{2019 IEEE/ACM 41st International Conference on Software
  Engineering: Software Engineering in Practice}, ICSE-SEIP 2019, pages
  291--300, Washington, DC, 2019. IEEE.
\newblock \doi{10.1109/ICSE-SEIP.2019.00042}.

\bibitem[Anderson and Anderson(2007)]{Anderson2007}
Michael Anderson and Susan~Leigh Anderson.
\newblock {Machine Ethics: Creating an Ethical Intelligent Agent}.
\newblock \emph{AI Mag.}, 28\penalty0 (4):\penalty0 15--26, 2007.
\newblock \doi{10.1609/AIMAG.v28i4.2065}.

\bibitem[Andrews et~al.(2024)Andrews, Smart, and Birhane]{Andrews2024}
Mel Andrews, Andrew Smart, and Abeba Birhane.
\newblock {The Reanimation of Pseudoscience in Machine Learning and Its Ethical
  Repercussions}.
\newblock \emph{Patterns}, 5\penalty0 (9), 2024.
\newblock \doi{10.1016/J.PATTER.2024.101027}.

\bibitem[Anil and Bao(2021)]{anil2021learning}
Cem Anil and Xuchan Bao.
\newblock {Learning to Elect}.
\newblock In \emph{Advances in Neural Information Processing Systems 34},
  NeurIPS 2021, pages 8006--8017, Red Hook, NY, 2021. Curran Associates.
\newblock URL
  \url{https://proceedings.neurips.cc/paper_files/paper/2021/hash/42d6c7d61481d1c21bd1635f59edae05-Abstract.html}.

\bibitem[Aroyo and Welty(2015)]{aroyo2015truth}
Lora Aroyo and Chris Welty.
\newblock {Truth is a Lie: Crowd Truth and the Seven Myths of Human
  Annotation}.
\newblock \emph{AI Mag.}, 36\penalty0 (1):\penalty0 15--24, 2015.
\newblock \doi{10.1609/AIMAG.v36i1.2564}.

\bibitem[Aydin et~al.(2014)Aydin, Yilmaz, Li, Li, Gao, and
  Demirbas]{aydin2014crowdsourcing}
Bahadir~Ismail Aydin, Yavuz~Selim Yilmaz, Yaliang Li, Qi~Li, Jing Gao, and
  Murat Demirbas.
\newblock {Crowdsourcing for Multiple-Choice Question Answering}.
\newblock \emph{Proc. AAAI Conf. Artif. Intell.}, 28\penalty0 (2):\penalty0
  2946--2953, 2014.
\newblock \doi{10.1609/AAAI.v28i2.19016}.

\bibitem[Aytaç(2024)]{Aytaç2024}
Uğur Aytaç.
\newblock {Big Tech, Algorithmic Power, and Democratic Control}.
\newblock \emph{J. Politics}, 86\penalty0 (4):\penalty0 1431--1445, 2024.
\newblock \doi{10.1086/729938}.

\bibitem[Bahrami(2025)]{Bahrami2025}
Nasser Bahrami.
\newblock {\emph{AI}gemony: Power Dynamics, Dominant Narratives, and
  Colonisation}.
\newblock \emph{AI Ethics}, 5\penalty0 (5):\penalty0 5081--5103, 2025.
\newblock \doi{10.1007/S43681-025-00734-4}.

\bibitem[Bai et~al.(2022)Bai, Kadavath, Kundu, Askell, Kernion, Jones, Chen,
  Goldie, Mirhoseini, McKinnon, Chen, Olsson, Olah, Hernandez, Drain, Ganguli,
  Li, Tran-Johnson, Perez, Kerr, Mueller, Ladish, Landau, Ndousse, Lukosuite,
  Lovitt, Sellitto, Elhage, Schiefer, Mercado, DasSarma, Lasenby, Larson,
  Ringer, Johnston, Kravec, Showk, Fort, Lanham, Telleen-Lawton, Conerly,
  Henighan, Hume, Bowman, Hatfield-Dodds, Mann, Amodei, Joseph, McCandlish,
  Brown, and Kaplan]{Bai2022}
Yuntao Bai, Saurav Kadavath, Sandipan Kundu, Amanda Askell, Jackson Kernion,
  Andy Jones, Anna Chen, Anna Goldie, Azalia Mirhoseini, Cameron McKinnon,
  Carol Chen, Catherine Olsson, Christopher Olah, Danny Hernandez, Dawn Drain,
  Deep Ganguli, Dustin Li, Eli Tran-Johnson, Ethan Perez, Jamie Kerr, Jared
  Mueller, Jeffrey Ladish, Joshua Landau, Kamal Ndousse, Kamile Lukosuite,
  Liane Lovitt, Michael Sellitto, Nelson Elhage, Nicholas Schiefer, Noemi
  Mercado, Nova DasSarma, Robert Lasenby, Robin Larson, Sam Ringer, Scott
  Johnston, Shauna Kravec, Sheer~El Showk, Stanislav Fort, Tamera Lanham,
  Timothy Telleen-Lawton, Tom Conerly, Tom Henighan, Tristan Hume, Samuel~R.
  Bowman, Zac Hatfield-Dodds, Ben Mann, Dario Amodei, Nicholas Joseph, Sam
  McCandlish, Tom Brown, and Jared Kaplan.
\newblock {Constitutional AI: Harmlessness from AI Feedback}, 2022.

\bibitem[Berengueres(2024)]{Berengueres2024}
Jose Berengueres.
\newblock {How to Regulate Large Language Models for Responsible AI}.
\newblock \emph{IEEE Trans. Technol. Soc.}, 5\penalty0 (2):\penalty0 191--197,
  2024.
\newblock \doi{10.1109/TTS.2024.3403681}.

\bibitem[Bogiatzis-Gibbons(2024)]{BogiatzisGibbons2024}
Daniel~James Bogiatzis-Gibbons.
\newblock {Beyond Individual Accountability: (Re-)Asserting Democratic Control
  of AI}.
\newblock In \emph{Proceedings of the 2024 ACM Conference on Fairness,
  Accountability, and Transparency}, FAccT '24, pages 74--84, New York, NY,
  2024. ACM.
\newblock \doi{10.1145/3630106.3658541}.

\bibitem[Bogomolnaia et~al.(2005)Bogomolnaia, Moulin, and
  Stong]{bogomolnaia2004dichotomousSDS}
Anna Bogomolnaia, Hervé Moulin, and Richard Stong.
\newblock {Collective Choice under Dichotomous Preferences}.
\newblock \emph{J. Econ. Theory}, 122\penalty0 (2):\penalty0 165--184, 2005.
\newblock \doi{10.1016/J.JET.2004.05.005}.

\bibitem[Brandt et~al.(2016)Brandt, Conitzer, Endriss, Lang, and
  Procaccia]{brandt2016handbook}
Felix Brandt, Vincent Conitzer, Ulle Endriss, Jérôme Lang, and Ariel
  Procaccia, editors.
\newblock \emph{{Handbook of Computational Social Choice}}.
\newblock Cambridge University Press, Cambridge, 2016.
\newblock ISBN 978-1-107-06043-2.

\bibitem[Braylan and Lease(2021)]{braylan2021aggregate_merge_match}
Alexander Braylan and Matthew Lease.
\newblock {Aggregating Complex Annotations via Merging and Matching}.
\newblock In \emph{Proceedings of the 27th ACM SIGKDD Conference on Knowledge
  Discovery and Data Mining}, KDD '21, pages 86--94, New York, NY, 2021. ACM.
\newblock \doi{10.1145/3447548.3467411}.

\bibitem[Bredereck et~al.(2024)Bredereck, Faliszewski, Igarashi, Lackner, and
  Skowron]{aaai/BFILS-diversity}
Robert Bredereck, Piotr Faliszewski, Ayumi Igarashi, Martin Lackner, and Piotr
  Skowron.
\newblock {Multiwinner Elections with Diversity Constraints}.
\newblock \emph{Proc. AAAI Conf. Artif. Intell.}, 32\penalty0 (1):\penalty0
  933--940, 2024.
\newblock \doi{10.1609/AAAI.v32i1.11457}.

\bibitem[Burka et~al.(2022)Burka, Puppe, Szepesv{\'a}ry, and
  Tasn{\'a}di]{burka2022voting}
D{\'a}vid Burka, Clemens Puppe, L{\'a}szl{\'o} Szepesv{\'a}ry, and Attila
  Tasn{\'a}di.
\newblock {Voting: A Machine Learning Approach}.
\newblock \emph{Eur. J. Oper. Res.}, 299\penalty0 (3):\penalty0 1003--1017,
  2022.
\newblock \doi{10.1016/J.EJOR.2021.10.005}.

\bibitem[Burke and Klein(2020)]{Burke2020bayesianinference}
Pierce Burke and Richard Klein.
\newblock {Confident in the Crowd: Bayesian Inference to Improve Data Labelling
  in Crowdsourcing}.
\newblock In \emph{2020 International SAUPEC/RobMech/PRASA Conference},
  SAUPEC/RobMech/PRASA 2020, Washington, DC, 2020. IEEE.
\newblock \doi{10.1109/SAUPEC/RobMech/PRASA48453.2020.9041099}.

\bibitem[Böhmer and Niedermeier(2021)]{BoehmerN21}
Niclas Böhmer and Rolf Niedermeier.
\newblock {Broadening the Research Agenda for Computational Social Choice:
  Multiple Preference Profiles and Multiple Solutions}.
\newblock In \emph{Proceedings of the 20th International Conference on
  Autonomous Agents and Multiagent Systems}, AAMAS '21, Richland, SC, 2021.
  IFAAMAS.
\newblock URL \url{https://www.ifaamas.org/Proceedings/aamas2021/pdfs/p1.pdf}.

\bibitem[Casper et~al.(2023)Casper, Davies, Shi, {Krendl Gilbert}, Scheurer,
  Rando, Freedman, Korbak, Lindner, Freire, Wang, Marks, Ségerie, Carroll,
  Peng, Christoffersen, Damani, Slocum, Anwar, Siththaranjan, Nadeau, Michaud,
  Pfau, Krasheninnikov, Chen, Langosco, Hase, Bıyık, Dragan, Krueger, Sadigh,
  and Hadfield-Menell]{Casper2023RLHF}
Stephen Casper, Xander Davies, Claudia Shi, Thomas {Krendl Gilbert}, Jérémy
  Scheurer, Javier Rando, Rachel Freedman, Tomasz Korbak, David Lindner, Pedro
  Freire, Tony Wang, Samuel Marks, Charbel-Raphaël Ségerie, Micah Carroll,
  Andi Peng, Phillip Christoffersen, Mehul Damani, Stewart Slocum, Usman Anwar,
  Anand Siththaranjan, Max Nadeau, Eric~J. Michaud, Jacob Pfau, Dmitrii
  Krasheninnikov, Xin Chen, Lauro Langosco, Peter Hase, Erdem Bıyık, Anca
  Dragan, David Krueger, Dorsa Sadigh, and Dylan Hadfield-Menell.
\newblock {Open Problems and Fundamental Limitations of Reinforcement Learning
  from Human Feedback}.
\newblock \emph{Trans. Mach. Learn. Res.}, 2023.
\newblock URL \url{https://openreview.net/forum?id=bx24KpJ4Eb}.

\bibitem[Chaffer et~al.(2024)Chaffer, {von Goins II}, Okusanya, Cotlage, and
  Goldston]{chaffer2025decentralizedgovernance}
Tomer~Jordi Chaffer, Charles {von Goins II}, Bayo Okusanya, Dontrail Cotlage,
  and Justin Goldston.
\newblock {Decentralized Governance of Autonomous AI Agents}, 2024.

\bibitem[{Collective Intelligence Project}(2023)]{CIP2023}
{Collective Intelligence Project}.
\newblock {The Collective Intelligence Project}, 2023.
\newblock URL \url{https://www.cip.org/whitepaper}.

\bibitem[Conitzer et~al.(2024)Conitzer, Freedman, Heitzig, Holliday, Jacobs,
  Lambert, Mosse, Pacuit, Russell, Schoelkopf, Tewolde, and
  Zwicker]{Conitzer2024}
Vincent Conitzer, Rachel Freedman, Jobst Heitzig, Wesley~H. Holliday, Bob~M.
  Jacobs, Nathan Lambert, Milan Mosse, Eric Pacuit, Stuart Russell, Hailey
  Schoelkopf, Emanuel Tewolde, and William~S. Zwicker.
\newblock {Position: Social Choice Should Guide AI Alignment in Dealing with
  Diverse Human Feedback}.
\newblock In \emph{Proceedings of the 41st International Conference on Machine
  Learning}, ICML 2024/Proceedings of Machine Learning Research 235, pages
  9346--9360, Cambridge, MA, 2024. JMLR and Microtome Publishing.
\newblock URL \url{https://proceedings.mlr.press/v235/conitzer24a.html}.

\bibitem[Corbett-Davies et~al.(2023)Corbett-Davies, Gaebler, Nilforoshan,
  Shroff, and Goel]{CorbettDavies2023}
Sam Corbett-Davies, Johann~D. Gaebler, Hamed Nilforoshan, Ravi Shroff, and
  Sharad Goel.
\newblock {The Measure and Mismeasure of Fairness}.
\newblock \emph{J. Mach. Learn. Res.}, 24\penalty0 (312):\penalty0 1--117,
  2023.
\newblock URL \url{http://jmlr.org/papers/v24/22-1511.html}.

\bibitem[Costanza-Chock et~al.(2022)Costanza-Chock, Raji, and
  Buolamwini]{CostanzaChock2022}
Sasha Costanza-Chock, Inioluwa~Deborah Raji, and Joy Buolamwini.
\newblock {Who Audits the Auditors? Recommendations from a Field Scan of the
  Algorithmic Auditing Ecosystem}.
\newblock In \emph{Proceedings of the 2022 ACM Conference on Fairness,
  Accountability, and Transparency}, FAccT '22, pages 1571--1583, New York, NY,
  2022. ACM.
\newblock \doi{10.1145/3531146.3533213}.

\bibitem[{del Rey} et~al.(2025){del Rey}, Cruz, Franch, and
  Martínez-Fernández]{delrey2025estimatingdeeplearningenergy}
Santiago {del Rey}, Luís Cruz, Xavier Franch, and Silverio
  Martínez-Fernández.
\newblock {Estimating Deep Learning Energy Consumption Based on Model
  Architecture and Training Environment}, 2025.

\bibitem[Elkind et~al.(2023)Elkind, Suksompong, and
  Teh]{elkind2024portioningcardinal}
Edith Elkind, Warut Suksompong, and Nicholas Teh.
\newblock {Settling the Score: Portioning with Cardinal Preferences}.
\newblock In \emph{26th European Conference on Artificial Intelligence}, ECAI
  2023, Frontiers in Artificial Intelligence and Applications 372, pages
  621--628, Amsterdam, 2023. IOS Press.
\newblock \doi{10.3233/FAIA230324}.

\bibitem[Elkind et~al.(2024)Elkind, Obraztsova, and Teh]{ElkindOT24}
Edith Elkind, Svetlana Obraztsova, and Nicholas Teh.
\newblock {Temporal Fairness in Multiwinner Voting}.
\newblock \emph{Proc. AAAI Conf. Artif. Intell.}, 38\penalty0 (20):\penalty0
  22633--22640, 2024.
\newblock \doi{10.1609/AAAI.v38i20.30273}.

\bibitem[Elkind et~al.(2025{\natexlab{a}})Elkind, Neoh, and
  Teh]{elkind2025temporalchores}
Edith Elkind, Tzeh~Yuan Neoh, and Nicholas Teh.
\newblock {Not in My Backyard! Temporal Voting over Public Chores}.
\newblock In \emph{Proceedings of the Thirty-Fourth International Joint
  Conference on Artificial Intelligence}, IJCAI-25, pages 3814--3820. IJCAI,
  2025{\natexlab{a}}.
\newblock \doi{10.24963/IJCAI.2025/424}.

\bibitem[Elkind et~al.(2025{\natexlab{b}})Elkind, Obraztsova, Peters, and
  Teh]{Elkind2025temporalproportionality}
Edith Elkind, Svetlana Obraztsova, Jannik Peters, and Nicholas Teh.
\newblock {Verifying Proportionality in Temporal Voting}.
\newblock \emph{Proc. AAAI Conf. Artif. Intell.}, 39\penalty0 (13):\penalty0
  13805--13813, 2025{\natexlab{b}}.
\newblock \doi{10.1609/AAAI.v39i13.33509}.

\bibitem[Faliszewski et~al.(2017)Faliszewski, Skowron, Slinko, and
  Talmon]{faliszewski2017multiwinner}
Piotr Faliszewski, Piotr Skowron, Arkadii Slinko, and Nimrod Talmon.
\newblock {Multiwinner Voting: A New Challenge for Social Choice Theory}.
\newblock In Ulle Endriss, editor, \emph{Trends in Computational Social
  Choice}, chapter~2, pages 27--47. AI Access Foundation, El Segundo, CA, 2017.
\newblock URL
  \url{https://archive.illc.uva.nl/COST-IC1205/BookDocs/Chapters/TrendsCOMSOC-02.pdf}.

\bibitem[Freeman et~al.(2017)Freeman, Zahedi, and
  Conitzer]{freeman2017fairscdynamic}
Rupert Freeman, Seyed~Majid Zahedi, and Vincent Conitzer.
\newblock {Fair and Efficient Social Choice in Dynamic Settings}.
\newblock In \emph{Proceedings of the Twenty-Sixth International Joint
  Conference on Artificial Intelligence}, IJCAI-17, pages 4580--4587. IJCAI,
  2017.
\newblock \doi{10.24963/IJCAI.2017/639}.

\bibitem[Freeman et~al.(2021)Freeman, Pennock, Peters, and {Wortman
  Vaughan}]{FREEMAN2021105234}
Rupert Freeman, David~M. Pennock, Dominik Peters, and Jennifer {Wortman
  Vaughan}.
\newblock {Truthful Aggregation of Budget Proposals}.
\newblock \emph{J. Econ. Theory}, 193, 2021.
\newblock \doi{10.1016/J.JET.2021.105234}.

\bibitem[Friedgut et~al.(2011)Friedgut, Kalai, Keller, and
  Nisan]{friedgut2011quantitative}
Ehud Friedgut, Gil Kalai, Nathan Keller, and Noam Nisan.
\newblock {A Quantitative Version of the Gibbard--Satterthwaite Theorem for
  Three Alternatives}.
\newblock \emph{SIAM J. Comput.}, 40\penalty0 (3):\penalty0 934--952, 2011.
\newblock \doi{10.1137/090756740}.

\bibitem[Furendal(2025)]{Furendal2025}
Markus Furendal.
\newblock {Collective Ownership of AI}.
\newblock In Martin Hähnel and Regina Müller, editors, \emph{A Companion to
  Applied Philosophy of AI}, chapter~26, pages 372--386. John Wiley \& Sons,
  Hoboken, NJ, 2025.
\newblock \doi{10.1002/9781394238651.ch26}.

\bibitem[Gabriel(2020)]{Gabriel2020}
Iason Gabriel.
\newblock {Artificial Intelligence, Values, and Alignment}.
\newblock \emph{Minds Mach.}, 30\penalty0 (3):\penalty0 411--437, 2020.
\newblock \doi{10.1007/S11023-020-09539-2}.

\bibitem[Ge et~al.(2024)Ge, Halpern, Micha, Procaccia, Shapira, Vorobeychik,
  and Wu]{ge2024axioms}
Luise Ge, Daniel Halpern, Evi Micha, Ariel~D. Procaccia, Itai Shapira, Yevgeniy
  Vorobeychik, and Junlin Wu.
\newblock {Axioms for AI Alignment from Human Feedback}.
\newblock In \emph{Advances in Neural Information Processing Systems 37},
  NeurIPS 2024, pages 80439--80465, Red Hook, NY, 2024. Curran Associates.
\newblock \doi{10.52202/079017-2557}.

\bibitem[Gehrlein and Lepelley(2011)]{gehrlein2010voting}
William~V. Gehrlein and Dominique Lepelley.
\newblock \emph{{Voting Paradoxes and Group Coherence: The Condorcet Efficiency
  of Voting Rules}}.
\newblock Studies in Choice and Welfare. Springer, Berlin/Heidelberg, 2011.
\newblock ISBN 978-3-642-03106-9.
\newblock \doi{10.1007/978-3-642-03107-6}.

\bibitem[Gershman(2025)]{gershman2025subjectivefunctions}
Samuel~J. Gershman.
\newblock {Subjective Functions}, 2025.

\bibitem[Goanta et~al.(2023)Goanta, Aletras, Chalkidis, Ranchordás, and
  Spanakis]{Goanta2023}
Catalina Goanta, Nikolaos Aletras, Ilias Chalkidis, Sofia Ranchordás, and
  Gerasimos Spanakis.
\newblock {Regulation and NLP (RegNLP): Taming Large Language Models}.
\newblock In \emph{Proceedings of the 2023 Conference on Empirical Methods in
  Natural Language Processing}, EMNLP 2023, pages 8712--8724, Stroudsburg, PA,
  2023. ACL.
\newblock \doi{10.18653/V1/2023.EMNLP-MAIN.539}.

\bibitem[Hacker et~al.(2023)Hacker, Engel, and Mauer]{Hacker2023}
Philipp Hacker, Andreas Engel, and Marco Mauer.
\newblock {Regulating ChatGPT and other Large Generative AI Models}.
\newblock In \emph{Proceedings of the 2023 ACM Conference on Fairness,
  Accountability, and Transparency}, FAccT '23, pages 1112--1123, New York, NY,
  2023. ACM.
\newblock \doi{10.1145/3593013.3594067}.

\bibitem[Hagendorff(2022)]{hagendorff21}
Thilo Hagendorff.
\newblock {Blind Spots in AI Ethics}.
\newblock \emph{AI Ethics}, 2\penalty0 (4):\penalty0 851--867, 2022.
\newblock \doi{10.1007/S43681-021-00122-8}.

\bibitem[Halpern et~al.(2023)Halpern, Kehne, Procaccia, Tucker-Foltz, and
  W{\"u}thrich]{halpern2023representation}
Daniel Halpern, Gregory Kehne, Ariel~D. Procaccia, Jamie Tucker-Foltz, and
  Manuel W{\"u}thrich.
\newblock {Representation with Incomplete Votes}.
\newblock \emph{Proc. AAAI Conf. Artif. Intell.}, 37\penalty0 (5):\penalty0
  5657--5664, 2023.
\newblock \doi{10.1609/AAAI.v37i5.25702}.

\bibitem[{High-Level Expert Group on Artificial Intelligence}(2019)]{HLEG2019}
{High-Level Expert Group on Artificial Intelligence}.
\newblock {Ethics Guidelines for Trustworthy AI}, 2019.
\newblock URL
  \url{https://digital-strategy.ec.europa.eu/en/library/ethics-guidelines-trustworthy-ai}.

\bibitem[Hornischer and Terzopoulou(2025)]{hornischer2025learning}
Levin Hornischer and Zoi Terzopoulou.
\newblock {Learning How to Vote with Principles: Axiomatic Insights Into the
  Collective Decisions of Neural Networks}.
\newblock \emph{J. Artif. Intell. Res.}, 83, 2025.
\newblock \doi{10.1613/JAIR.1.18890}.

\bibitem[Huang et~al.(2024)Huang, Siddarth, Lovitt, Liao, Durmus, Tamkin, and
  Ganguli]{Anthropic2024}
Saffron Huang, Divya Siddarth, Liane Lovitt, Thomas~I. Liao, Esin Durmus, Alex
  Tamkin, and Deep Ganguli.
\newblock {Collective Constitutional AI: Aligning a Language Model with Public
  Input}.
\newblock In \emph{Proceedings of the 2024 ACM Conference on Fairness,
  Accountability, and Transparency}, FAccT '24, pages 1395--1417, New York, NY,
  2024. ACM.
\newblock \doi{10.1145/3630106.3658979}.

\bibitem[Indykov et~al.(2025)Indykov, Strüber, and
  Wohlrab]{INDYKOV2025architecturaltactics}
Vladislav Indykov, Daniel Strüber, and Rebekka Wohlrab.
\newblock {Architectural Tactics to Achieve Quality Attributes of
  Machine-Learning-Enabled Systems: A Systematic Literature Review}.
\newblock \emph{J. Syst. Softw.}, 223, 2025.
\newblock \doi{10.1016/J.JSS.2025.112373}.

\bibitem[Ji et~al.(2026)Ji, Qiu, Chen, Zhou, Zhang, Hong, Lou, Wang, Duan, He,
  Vierling, Zhang, Zeng, Dai, Pan, Xu, O'Gara, Ng, Tse, Fu, Mcaleer, Wang,
  Yang, Liu, Wang, Zhu, Guo, Yang, and Gao]{Ji2025alignment}
Jiaming Ji, Tianyi Qiu, Boyuan Chen, Jiayi Zhou, Borong Zhang, Donghai Hong,
  Hantao Lou, Kaile Wang, Yawen Duan, Zhonghao He, Lukas Vierling, Zhaowei
  Zhang, Fanzhi Zeng, Juntao Dai, Xuehai Pan, Hua Xu, Aidan O'Gara, Kwan Ng,
  Brian Tse, Jie Fu, Stephen Mcaleer, Yanfeng Wang, Mingchuan Yang, Yunhuai
  Liu, Yizhou Wang, Song-Chun Zhu, Yike Guo, Yaodong Yang, and Wen Gao.
\newblock {AI Alignment: A Contemporary Survey}.
\newblock \emph{ACM Comput. Surv.}, 58\penalty0 (5), 2026.
\newblock \doi{10.1145/3770749}.

\bibitem[Kallina et~al.(2025)Kallina, Bohn\'{e}, and Singh]{Kallina2025}
Emma Kallina, Thomas Bohn\'{e}, and Jatinder Singh.
\newblock {Stakeholder Participation for Responsible AI Development:
  Disconnects Between Guidance and Current Practice}.
\newblock In \emph{Proceedings of the 2025 ACM Conference on Fairness,
  Accountability, and Transparency}, FAccT '25, pages 1060--1079, New York, NY,
  2025. ACM.
\newblock \doi{10.1145/3715275.3732069}.

\bibitem[Kasy(2024)]{Kasy2024}
Maximilian Kasy.
\newblock {The Political Economy of AI: Towards Democratic Control of the Means
  of Prediction}.
\newblock Iza discussion papers, no.\ 16948, Institute of Labor Economics
  (IZA), Bonn, 2024.
\newblock URL \url{https://hdl.handle.net/10419/295971}.

\bibitem[Kasy(2025)]{Kasy2025}
Maximilian Kasy.
\newblock \emph{{The Means of Prediction: How AI Really Works (and Who
  Benefits)}}.
\newblock University of Chicago Press, Hoboken, NJ, 2025.
\newblock ISBN 978-0-226-83953-0.

\bibitem[Lackner(2020)]{aaai/perpetual}
Martin Lackner.
\newblock {Perpetual Voting: Fairness in Long-Term Decision Making}.
\newblock \emph{Proc. AAAI Conf. Artif. Intell.}, 34\penalty0 (2):\penalty0
  2103--2110, 2020.
\newblock \doi{10.1609/AAAI.v34i02.5584}.

\bibitem[Lackner and Maly(2023)]{aaai/proportional-perpetual-voting}
Martin Lackner and Jan Maly.
\newblock {Proportional Decisions in Perpetual Voting}.
\newblock \emph{Proc. AAAI Conf. Artif. Intell.}, 37\penalty0 (5):\penalty0
  5722--5729, 2023.
\newblock \doi{10.1609/AAAI.v37i5.25710}.

\bibitem[Lackner and Skowron(2023)]{Lackner22abcvoting}
Martin Lackner and Piotr Skowron.
\newblock \emph{{Multi-Winner Voting with Approval Preferences}}.
\newblock SpringerBriefs in Intelligent Systems. Springer, Cham, 2023.
\newblock ISBN 978-3-031-09015-8.

\bibitem[Lakoff(1987)]{lakoff1987categoryperception}
George Lakoff.
\newblock \emph{{Women, Fire, and Dangerous Things: What Categories Reveal
  about the Mind}}.
\newblock University of Chicago Press, Chicago, IL, 1987.
\newblock ISBN 0-226-46803-8.

\bibitem[List(2022)]{list-sc}
Christian List.
\newblock {Social Choice Theory}.
\newblock In Edward~N. Zalta and Uri Nodelman, editors, \emph{Stanford
  Encyclopedia of Philosophy}. Metaphysics Research Lab, Stanford University,
  Stanford, CA, 2022.
\newblock URL
  \url{https://plato.stanford.edu/archives/win2022/entries/social-choice/}.

\bibitem[Lowe et~al.(2025)Lowe, Edelman, Zhi-Xuan, Klingefjord, Hain, Wang,
  Sarkar, Bakker, Barez, Franklin, Haupt, Heitzig, Holliday, Jara-Ettinger,
  Kasirzadeh, Kearns, Kirkpatrick, Koh, Lehman, Levine, Revel, and
  Vendrov]{Lowe2025}
Ryan Lowe, Joe Edelman, Tan Zhi-Xuan, Oliver Klingefjord, Ellie Hain, Vincent
  Wang, Atrisha Sarkar, Michiel~A. Bakker, Fazl Barez, Matija Franklin, Andreas
  Haupt, Jobst Heitzig, Wesley~H. Holliday, Julian Jara-Ettinger, Atoosa
  Kasirzadeh, Ryan~Othniel Kearns, James~Ravi Kirkpatrick, Andrew Koh, Joel
  Lehman, Sydney Levine, Manon Revel, and Ivan Vendrov.
\newblock {Full-Stack Alignment: Co-Aligning AI and Institutions with Thicker
  Models of Value}.
\newblock In \emph{Proceedings of the 2nd Workshop on Models of Human Feedback
  for AI Alignment at the 42nd International Conference on Machine Learning},
  MoFA 2025, Cambridge, MA, 2025. JMLR and Microtome Publishing.

\bibitem[Lu et~al.(2019)Lu, Liu, Dong, Gu, Gama, and Zhang]{Lu2019}
Jie Lu, Anjin Liu, Fan Dong, Feng Gu, Jo{\~a}o Gama, and Guangquan Zhang.
\newblock {Learning under Concept Drift: A Review}.
\newblock \emph{IEEE Trans. Knowl. Data Eng.}, 31\penalty0 (12):\penalty0
  2346--2363, 2019.
\newblock \doi{10.1109/TKDE.2018.2876857}.

\bibitem[Lu and Boutilier(2020)]{lu20elicitation_multiwinner}
Tyler Lu and Craig Boutilier.
\newblock {Preference Elicitation and Robust Winner Determination for Single-
  and Multi-Winner Social Choice}.
\newblock \emph{Artif. Intell.}, 279, 2020.
\newblock \doi{10.1016/J.ARTINT.2019.103203}.

\bibitem[Makridakis et~al.(2023)Makridakis, Petropoulos, and
  Kang]{Makridakis2023}
Spyros Makridakis, Fotios Petropoulos, and Yanfei Kang.
\newblock {Large Language Models: Their Success and Impact}.
\newblock \emph{Forecast.}, 5\penalty0 (3):\penalty0 536--549, 2023.
\newblock \doi{10.3390/FORECAST5030030}.

\bibitem[Mandal et~al.(2024)Mandal, Lin, and Srikant]{Mandal2024}
Saptarshi Mandal, Xiaojun Lin, and Rayadurgam Srikant.
\newblock {A Theoretical Analysis of Soft-Label vs Hard-Label Training in
  Neural Networks}, 2024.

\bibitem[Matone et~al.(2024)Matone, Abramowitz, Armstrong, Balakrishnan, and
  Mattei]{matone2024deepvoting}
Leonardo Matone, Ben Abramowitz, Ben Armstrong, Avinash Balakrishnan, and
  Nicholas Mattei.
\newblock {DeepVoting: Learning and Fine-Tuning Voting Rules with Canonical
  Embeddings}, 2024.

\bibitem[Meskó and Topol(2023)]{Mesko2023}
Bertalan Meskó and Eric~J. Topol.
\newblock {The Imperative for Regulatory Oversight of Large Language Models (or
  Generative AI) in Healthcare}.
\newblock \emph{npj Digit. Med.}, 6, 2023.
\newblock \doi{10.1038/S41746-023-00873-0}.

\bibitem[{Metagov}(2025)]{metagov-cg-ai}
{Metagov}.
\newblock {Collective Governance for AI: Points of Intervention}, 2025.
\newblock URL \url{https://metagov.org/cg-ai/}.

\bibitem[Mohammed et~al.(2025)Mohammed, Budach, Feuerpfeil, Ihde, Nathansen,
  Noack, Patzlaff, Naumann, and Harmouch]{MOHAMMED2025dataquality}
Sedir Mohammed, Lukas Budach, Moritz Feuerpfeil, Nina Ihde, Andrea Nathansen,
  Nele Noack, Hendrik Patzlaff, Felix Naumann, and Hazar Harmouch.
\newblock {The Effects of Data Quality on Machine Learning Performance on
  Tabular Data}.
\newblock \emph{Inf. Syst.}, 132, 2025.
\newblock \doi{10.1016/J.IS.2025.102549}.

\bibitem[Mohsin et~al.(2022)Mohsin, Liu, Chen, Rossi, and
  Xia]{mohsin2022learning}
Farhad Mohsin, Ao~Liu, Pin-Yu Chen, Francesca Rossi, and Lirong Xia.
\newblock {Learning to Design Fair and Private Voting Rules}.
\newblock \emph{J. Artif. Intell. Res.}, 75:\penalty0 1139--1176, 2022.
\newblock \doi{10.1613/JAIR.1.13734}.

\bibitem[Moulin(1980)]{moulin1980strategy}
Herv{\'e} Moulin.
\newblock {On Strategy-Proofness and Single Peakedness}.
\newblock \emph{Public Choice}, 35\penalty0 (4):\penalty0 437--455, 1980.
\newblock \doi{10.1007/BF00128122}.

\bibitem[Moulin(1981)]{moulin1981proportional}
Herv{\'e} Moulin.
\newblock {The Proportional Veto Principle}.
\newblock \emph{Rev. Econ. Stud.}, 48\penalty0 (3):\penalty0 407--416, 1981.
\newblock \doi{10.2307/2297154}.

\bibitem[Moulin(1988{\natexlab{a}})]{moulin1988condorcet}
Herv{\'e} Moulin.
\newblock {Condorcet's Principle Implies the No Show Paradox}.
\newblock \emph{J. Econ. Theory}, 45\penalty0 (1):\penalty0 53--64,
  1988{\natexlab{a}}.
\newblock \doi{10.1016/0022-0531(88)90253-0}.

\bibitem[Moulin(1988{\natexlab{b}})]{Moul88a}
Hervé Moulin.
\newblock \emph{{Axioms of Cooperative Decision Making}}.
\newblock Cambridge University Press, Cambridge, 1988{\natexlab{b}}.
\newblock ISBN 978-0-521-36055-5.
\newblock \doi{10.1017/CCOL0521360552}.

\bibitem[Noothigattu et~al.(2018)Noothigattu, Gaikwad, Awad, Dsouza, Rahwan,
  Ravikumar, and Procaccia]{noothigattu2018voting}
Ritesh Noothigattu, Snehalkumar Gaikwad, Edmond Awad, Sohan Dsouza, Iyad
  Rahwan, Pradeep Ravikumar, and Ariel Procaccia.
\newblock {A Voting-Based System for Ethical Decision Making}.
\newblock \emph{Proc. AAAI Conf. Artif. Intell.}, 32\penalty0 (1):\penalty0
  1587--1594, 2018.
\newblock \doi{10.1609/AAAI.v32i1.11512}.

\bibitem[Ovadya et~al.(2025)Ovadya, Redman, Thorburn, Chen, Smith, Devine,
  Konya, Milli, Revel, Feng, Zhang, Chandra, Bakker, and
  Kasirzadeh]{Ovadya2025}
Aviv Ovadya, Kyle Redman, Luke Thorburn, Quan~Ze Chen, Oliver Smith, Flynn
  Devine, Andrew Konya, Smitha Milli, Manon Revel, Kevin Feng, Amy~X. Zhang,
  Bilva Chandra, Michiel~A. Bakker, and Atoosa Kasirzadeh.
\newblock {Position: Democratic AI is Possible. The Democracy Levels Framework
  Shows How It Might Work}.
\newblock In \emph{Proceedings of the 42nd International Conference on Machine
  Learning}, ICML 2025/Proceedings of Machine Learning Research 267, pages
  81930--81961, Cambridge, MA, 2025. JMLR and Microtome Publishing.
\newblock URL \url{https://proceedings.mlr.press/v267/ovadya25a.html}.

\bibitem[Parkes and Procaccia(2013)]{parkes2013dynamicscevolving}
David~C. Parkes and Ariel Procaccia.
\newblock {Dynamic Social Choice with Evolving Preferences}.
\newblock \emph{Proc. AAAI Conf. Artif. Intell.}, 27\penalty0 (1):\penalty0
  767--773, 2013.
\newblock \doi{10.1609/AAAI.v27i1.8570}.

\bibitem[Paullada et~al.(2021)Paullada, Raji, Bender, Denton, and
  Hanna]{paullada2021data}
Amandalynne Paullada, Inioluwa~Deborah Raji, Emily~M. Bender, Emily Denton, and
  Alex Hanna.
\newblock {Data and Its (Dis)Contents: A Survey of Dataset Development and Use
  in Machine Learning Research}.
\newblock \emph{Patterns}, 2\penalty0 (11), 2021.
\newblock \doi{10.1016/J.PATTER.2021.100336}.

\bibitem[Peterson et~al.(2019)Peterson, Battleday, Griffiths, and
  Russakovsky]{Peterson2019}
Joshua Peterson, Ruairidh Battleday, Thomas Griffiths, and Olga Russakovsky.
\newblock {Human Uncertainty Makes Classification More Robust}.
\newblock In \emph{2019 IEEE/CVF International Conference on Computer Vision},
  ICCV 2019, pages 9616--9625, Washington, DC, 2019. IEEE.
\newblock \doi{10.1109/ICCV.2019.00971}.

\bibitem[Pivato(2013)]{Pivato2013comsoc_mle}
Marcus Pivato.
\newblock {Voting Rules as Statistical Estimators}.
\newblock \emph{Soc. Choice Welf.}, 40\penalty0 (2):\penalty0 581--630, 2013.
\newblock \doi{10.1007/S00355-011-0619-1}.

\bibitem[Plank(2022)]{Plank22}
Barbara Plank.
\newblock {The “Problem” of Human Label Variation: On Ground Truth in Data,
  Modeling and Evaluation}.
\newblock In \emph{Proceedings of the 2022 Conference on Empirical Methods in
  Natural Language Processing}, EMNLP 2022, pages 10671--10682, Stroudsburg,
  PA, 2022. ACL.
\newblock \doi{10.18653/V1/2022.EMNLP-MAIN.731}.

\bibitem[Platanios et~al.(2020)Platanios, Al-Shedivat, Xing, and
  Mitchell]{platanios2020learningimperfectannotations}
Emmanouil~Antonios Platanios, Maruan Al-Shedivat, Eric Xing, and Tom Mitchell.
\newblock {Learning from Imperfect Annotations}, 2020.

\bibitem[Prainsack(2025)]{Prainsack2025}
Barbara Prainsack.
\newblock {Our Stakes in Data: How Do We (Re)Gain Democratic Control Over
  Digital Practices?}
\newblock In Jurgen Goossens, Esther Keymolen, and Antonia Stanojević,
  editors, \emph{Public Governance and Emerging Technologies: Values, Trust,
  and Regulatory Compliance}, pages 131--147. Springer International
  Publishing, Cham, 2025.
\newblock \doi{10.1007/978-3-031-84748-6_7}.

\bibitem[Procaccia et~al.(2009)Procaccia, Zohar, Peleg, and
  Rosenschein]{procaccia2009learnability}
Ariel~D. Procaccia, Aviv Zohar, Yoni Peleg, and Jeffrey~S. Rosenschein.
\newblock {The Learnability of Voting Rules}.
\newblock \emph{Artif. Intell.}, 173\penalty0 (12--13):\penalty0 1133--1149,
  2009.
\newblock \doi{10.1016/J.ARTINT.2009.03.003}.

\bibitem[Raji et~al.(2020)Raji, Smart, White, Mitchell, Gebru, Hutchinson,
  Smith-Loud, Theron, and Barnes]{Raji2020}
Inioluwa~Deborah Raji, Andrew Smart, Rebecca~N. White, Margaret Mitchell,
  Timnit Gebru, Ben Hutchinson, Jamila Smith-Loud, Daniel Theron, and Parker
  Barnes.
\newblock {Closing the AI Accountability Gap: Defining an End-to-End Framework
  for Internal Algorithmic Auditing}.
\newblock In \emph{Proceedings of the 2020 Conference on Fairness,
  Accountability, and Transparency}, FAT* '20, pages 33--44, New York, NY,
  2020. ACM.
\newblock \doi{10.1145/3351095.3372873}.

\bibitem[Reddi(2026)]{mlbook2025}
Vijay~Janapa Reddi.
\newblock \emph{{Introduction to Machine Learning Systems}}.
\newblock MIT Press, Cambridge, MA, 2026.
\newblock URL \url{https://mlsysbook.ai/}.

\bibitem[Regenwetter et~al.(2006)Regenwetter, Grofman, Marley, and
  Tsetlin]{regenwetter2006behavioral}
Michel Regenwetter, Bernard Grofman, A.~A.~J. Marley, and Ilia Tsetlin.
\newblock \emph{{Behavioral Social Choice: Probabilistic Models, Statistical
  Inference, and Applications}}.
\newblock Cambridge University Press, Cambridge, 2006.
\newblock ISBN 978-0-521-82968-7.

\bibitem[Rehak(2025)]{rehak25}
Rainer Rehak.
\newblock {Catastrophic Computation. On the Impossibility of Sustainable
  Artificial Intelligence}.
\newblock In \emph{Digital Humanism}, DIGHUM 2025, Lecture Notes in Computer
  Science 16319, pages 110--118, Cham, 2025. Springer.
\newblock \doi{10.1007/978-3-032-11108-1_8}.

\bibitem[Sangiovanni(2019)]{Sangiovanni2019}
Andrea Sangiovanni.
\newblock {Democratic Control of Information in the Age of Surveillance
  Capitalism}.
\newblock \emph{J. Appl. Philos.}, 36\penalty0 (2):\penalty0 212--216, 2019.
\newblock \doi{10.1111/JAPP.12363}.

\bibitem[Seger et~al.(2023)Seger, Ovadya, Siddarth, Garfinkel, and
  Dafoe]{seger2023DemocratisingAIMultiple}
Elizabeth Seger, Aviv Ovadya, Divya Siddarth, Ben Garfinkel, and Allan Dafoe.
\newblock {Democratising AI: Multiple Meanings, Goals, and Methods}.
\newblock In \emph{Proceedings of the 2023 AAAI/ACM Conference on AI, Ethics,
  and Society}, AIES '23, pages 715--722, New York, NY, 2023. ACM.
\newblock \doi{10.1145/3600211.3604693}.

\bibitem[Selbst et~al.(2019)Selbst, danah boyd, Friedler, Venkatasubramanian,
  and Vertesi]{SelbstBFVV19}
Andrew~D. Selbst, danah boyd, Sorelle~A. Friedler, Suresh Venkatasubramanian,
  and Janet Vertesi.
\newblock {Fairness and Abstraction in Sociotechnical Systems}.
\newblock In \emph{Proceedings of the 2019 Conference on Fairness,
  Accountability, and Transparency}, FAT* '19, pages 59--68, New York, NY,
  2019. ACM.
\newblock \doi{10.1145/3287560.3287598}.

\bibitem[Shah et~al.(2023)Shah, Entwistle, and Pfeffer]{Shah2023}
Nigam~H. Shah, David Entwistle, and Michael~A. Pfeffer.
\newblock {Creation and Adoption of Large Language Models in Medicine}.
\newblock \emph{JAMA}, 330\penalty0 (9):\penalty0 866--869, 2023.
\newblock \doi{10.1001/JAMA.2023.14217}.

\bibitem[Sloane et~al.(2022)Sloane, Moss, Awomolo, and
  Forlano]{sloane22participation}
Mona Sloane, Emanuel Moss, Olaitan Awomolo, and Laura Forlano.
\newblock {Participation Is Not a Design Fix for Machine Learning}.
\newblock In \emph{Proceedings of the 2nd ACM Conference on Equity and Access
  in Algorithms, Mechanisms, and Optimization}, EAAMO '22, New York, NY, 2022.
  ACM.
\newblock \doi{10.1145/3551624.3555285}.

\bibitem[Sorensen et~al.(2024)Sorensen, Moore, Fisher, Gordon, Mireshghallah,
  Rytting, Ye, Jiang, Lu, Dziri, Althoff, and
  Choi]{sorensen2024pluralisticalignment}
Taylor Sorensen, Jared Moore, Jillian Fisher, Mitchell Gordon, Niloofar
  Mireshghallah, Christopher~Michael Rytting, Andre Ye, Liwei Jiang, Ximing Lu,
  Nouha Dziri, Tim Althoff, and Yejin Choi.
\newblock {Position: A Roadmap to Pluralistic Alignment}.
\newblock In \emph{Proceedings of the 41st International Conference on Machine
  Learning}, ICML 2024/Proceedings of Machine Learning Research 235, pages
  46280--46302, Cambridge, MA, 2024. JMLR and Microtome Publishing.
\newblock URL \url{https://proceedings.mlr.press/v235/sorensen24a.html}.

\bibitem[Stark and Hutson(2022)]{Stark2022}
Luke Stark and Jevan Hutson.
\newblock {Physiognomic Artificial Intelligence}.
\newblock \emph{Fordham Intellect. Prop. Media Entertain. Law J.}, 32\penalty0
  (4):\penalty0 922--978, 2022.
\newblock URL \url{https://ir.lawnet.fordham.edu/iplj/vol32/iss4/2}.

\bibitem[Strubell et~al.(2019)Strubell, Ganesh, and
  McCallum]{strubell-etal-2019-energy}
Emma Strubell, Ananya Ganesh, and Andrew McCallum.
\newblock {Energy and Policy Considerations for Deep Learning in NLP}.
\newblock In \emph{Proceedings of the 57th Annual Meeting of the Association
  for Computational Linguistics}, ACL 2019, pages 3645--3650, Stroudsburg, PA,
  2019. ACL.
\newblock \doi{10.18653/v1/P19-1355}.

\bibitem[Suresh et~al.(2024)Suresh, Tseng, Young, Gray, Pierson, and
  Levy]{Suresh2024}
Harini Suresh, Emily Tseng, Meg Young, Mary Gray, Emma Pierson, and Karen Levy.
\newblock {Participation in the Age of Foundation Models}.
\newblock In \emph{Proceedings of the 2024 ACM Conference on Fairness,
  Accountability, and Transparency}, FAccT '24, pages 1609--1621, New York, NY,
  2024. ACM.
\newblock \doi{10.1145/3630106.3658992}.

\bibitem[Uma et~al.(2020)Uma, Fornaciari, Hovy, Paun, Plank, and
  Poesio]{Uma2020}
Alexandra Uma, Tommaso Fornaciari, Dirk Hovy, Silviu Paun, Barbara Plank, and
  Massimo Poesio.
\newblock {A Case for Soft Loss Functions}.
\newblock \emph{Proc. AAAI Conf. Hum. Comput. Crowdsourcing}, 8\penalty0
  (1):\penalty0 173--177, 2020.
\newblock \doi{10.1609/hcomp.v8i1.7478}.

\bibitem[Uma et~al.(2021)Uma, Fornaciari, Hovy, Paun, Plank, and
  Poesio]{uma2022disagreementlearning}
Alexandra~N. Uma, Tommaso Fornaciari, Dirk Hovy, Silviu Paun, Barbara Plank,
  and Massimo Poesio.
\newblock {Learning from Disagreement: A Survey}.
\newblock \emph{J. Artif. Intell. Res.}, 72:\penalty0 1385--1470, 2021.
\newblock \doi{10.1613/JAIR.1.12752}.

\bibitem[{Working Group on Artificial Intelligence and its Implications for the
  Information and Communication Space}(2024)]{WGAI2024}
{Working Group on Artificial Intelligence and its Implications for the
  Information and Communication Space}.
\newblock {AI as a Public Good: Ensuring Democratic Control of AI in the
  Information Space}, 2024.
\newblock URL
  \url{https://informationdemocracy.org/wp-content/uploads/2024/03/ID-AI-as-a-Public-Good-Feb-2024.pdf}.

\bibitem[Xia(2013)]{xia2013designing}
Lirong Xia.
\newblock {Designing Social Choice Mechanisms Using Machine Learning}.
\newblock In \emph{Proceedings of the 2013 International Conference on
  Autonomous Agents and Multi-Agent Systems}, AAMAS '13, pages 471--474,
  Richland, SC, 2013. IFAAMAS.
\newblock URL
  \url{https://www.ifaamas.org/Proceedings/aamas2013/docs/p471.pdf}.

\bibitem[Yan et~al.(2024)Yan, Sha, Zhao, Li, Martinez-Maldonado, Chen, Li, Jin,
  and Gašević]{Yan2024}
Lixiang Yan, Lele Sha, Linxuan Zhao, Yuheng Li, Roberto Martinez-Maldonado,
  Guanliang Chen, Xinyu Li, Yueqiao Jin, and Dragan Gašević.
\newblock {Practical and Ethical Challenges of Large Language Models in
  Education: A Systematic Scoping Review}.
\newblock \emph{Br. J. Educ. Technol.}, 55\penalty0 (1):\penalty0 90--112,
  2024.
\newblock \doi{10.1111/BJET.13370}.

\bibitem[Yin et~al.(2017)Yin, Han, Zhang, and
  Yu]{yin2017aggregatingcrowdwisdoms}
Li'ang Yin, Jianhua Han, Weinan Zhang, and Yong Yu.
\newblock {Aggregating Crowd Wisdoms with Label-aware Autoencoders}.
\newblock In \emph{Proceedings of the Twenty-Sixth International Joint
  Conference on Artificial Intelligence}, IJCAI-17, pages 1325--1331. IJCAI,
  2017.
\newblock \doi{10.24963/IJCAI.2017/184}.

\bibitem[Zhi-Xuan et~al.(2025)Zhi-Xuan, Carroll, Franklin, and
  Ashton]{ZhiXuan2025}
Tan Zhi-Xuan, Micah Carroll, Matija Franklin, and Hal Ashton.
\newblock {Beyond Preferences in AI Alignment}.
\newblock \emph{Philos. Stud.}, 182\penalty0 (7):\penalty0 1813--1863, 2025.
\newblock \doi{10.1007/S11098-024-02249-W}.

\bibitem[Zimmermann et~al.(2025)Zimmermann, Zeppa, Pandey, and
  Diao]{Zimmermann2025}
Annette Zimmermann, Andrew Zeppa, Srijan Pandey, and Kenneth Diao.
\newblock {Don't Give Up on Democratizing {AI} for the Wrong Reasons}.
\newblock In \emph{Advances in Neural Information Processing Systems 38},
  NeurIPS 2025, Red Hook, NY, 2025. Curran Associates.

\end{thebibliography}

\end{document}